\documentclass{article}
\usepackage[english]{babel}
\usepackage[utf8]{inputenc}
\usepackage{color}

\usepackage{arxiv}

\usepackage{graphicx} 
\usepackage{amsmath} 
\usepackage{amssymb} 
\usepackage{amsthm} 
\usepackage{tabularx} 
\usepackage{siunitx} 
\usepackage{algorithm}
\usepackage{float} 
\usepackage{caption} 
\usepackage{subcaption} 
\usepackage{filecontents} 
\usepackage{epstopdf}
\usepackage[makeroom]{cancel} 
\usepackage[noend]{algpseudocode}
\usepackage{gensymb} 
\usepackage{comment}

\usepackage[T1]{fontenc}    
\usepackage{lmodern}
\usepackage[colorlinks=true, urlcolor=black, linkcolor=blue, citecolor=red]{hyperref}
\usepackage{url}            
\usepackage{booktabs}       
\usepackage{amsfonts}       
\usepackage{nicefrac}       
\usepackage{lipsum}		
\usepackage{doi}

\usepackage{newclude}			

\usepackage[sc,osf]{mathpazo}   
\everymath{\displaystyle}
\graphicspath{{img/}{Figures/}}

\theoremstyle{remark}

\usepackage{stmaryrd}
\usepackage{subcaption}
\title{Simulation of continuous dynamic recrystallization using a level-set method}

\author{ \href{https://orcid.org/0000-0002-3382-4730}{\includegraphics[scale=0.06]{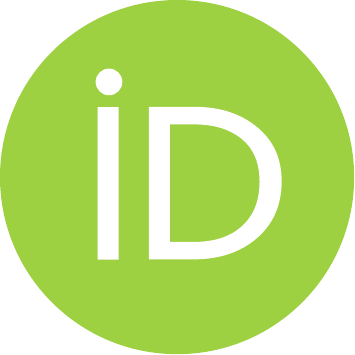}\hspace{1mm}Victor~Grand}\\
	Framatome - Components Research Center\\
	Mines Paris, PSL University\\
	Centre for material forming (CEMEF) \\
	UMR CNRS, 06904 Sophia Antipolis, France\\
	\texttt{victor.grand@framatome.com, victor.grand@minesparis.psl.eu} \\
	\And
	\href{https://orcid.org/0000-0001-6804-1974}{\includegraphics[scale=0.06]{orcid.pdf}\hspace{1mm}Baptiste~Flipon} \\
	Mines Paris, PSL University\\
	Centre for material forming (CEMEF) \\
	UMR CNRS, 06904 Sophia Antipolis, France\\
	\texttt{baptiste.flipon@minesparis.psl.eu} \\
	\And
         Alexis~Gaillac \\
	Framatome - Components Research Center\\
	France\\
	\texttt{alexis.gaillac@framatome.com} \\
	\And
	\href{https://orcid.org/0000-0002-6677-2850}{\includegraphics[scale=0.06]{orcid.pdf}\hspace{1mm}Marc~Bernacki} \\
	Mines Paris, PSL University\\
	Centre for material forming (CEMEF) \\
	UMR CNRS, 06904 Sophia Antipolis, France\\
	\texttt{marc.bernacki@minesparis.psl.eu} \\
}

\renewcommand{\shorttitle}{\textit{arXiv} Template}

\hypersetup{
pdftitle={Simulation of continuous dynamic recrystallization using a level-set method},
pdfsubject={},
pdfauthor={Victor~Grand, Baptiste~Flipon, Alexis~Gaillac, Marc~Bernacki},
pdfkeywords={Level-set, Grain growth, Continuous dynamic recrystallization,
Numerical simulation, Hot forming},
}

\begin{document}
\maketitle

\begin{abstract}
Dynamic recrystallization is one of the main phenomena responsible for microstructure evolutions during hot forming. Consequently, getting a better understanding of DRX mechanisms and being able to predict them is crucial. This paper proposes a full-field numerical framework to predict the evolution of subgrain structures upon grain growth, continuous dynamic and post-dynamic recrystallization. The microstructure representation into the numerical environment is presented. The developments made to improve substructure description are detailed extensively. Using these simulation tools, simulation of grain growth of a fully substructured microstructure are run. The influence of microstructure topology, of subgrain parameters and of some remaining stored energy due to plastic deformation is discussed. An analysis of the criterion for discrimination of recrystallized grains is proposed. Finally, the ability of the framework to model continuous dynamic and post-dynamic recrystallization is assessed upon a case study. The influence of grain boundary properties and of nucleation rules are studied. The representativity of the results in regards of experimental data will be discussed in an upcoming article.
\end{abstract}


\section{Introduction}\label{sec:Intro}

Dynamic recrystallization (DRX) is one of the main phenomena responsible for microstructure evolution during hot forming operations of metallic materials. Improving our ability to model those mechanisms is of first interest since it would allow to assess the influence of material and processing parameters and to reduce the number of experiments required to optimize industrial manufacturing paths \cite{Rollett2017}.
DRX is defined as the formation of new grains with low dislocation density that progressively consume the deformed microstructure under hot deformation conditions \cite{Rollett2017, Doherty1997}. However, this mechanism of microstructure evolution can exhibit various typical features. Therefore, based on those characteristics, DRX is commonly classified under three categories \cite{Huang2016}:
\begin{itemize}
\item discontinuous (DDRX) if recrystallized grains nucleate at some specific locations, generally close to grain boundaries (GB), and then grow and consume the deformed grains surrounding them. DDRX is therefore characterized by spatial and temporal discontinuity at the polycrystal scale.
\item continuous (CDRX) when recrystallized grains form slowly and continuously during deformation. In that case, grain formation is induced by the progressive reorganization of dislocations into cells or subgrains and the gradual increase of misorientation angle between those subgrains.
\item geometric (GDRX) at large strains when grains become serrated and some GB start to meet and enclose new grains.
\end{itemize}

It should be pointed out that this terminology is sometimes used inappropriately and could lead to confusion. CDRX and DDRX are based on the same physical phenomena that simply take place at different spatial and temporal scale which leads to those highly different observed features.
The predominance of one or the other mechanism is influenced by both material characteristics and thermomechanical conditions. DDRX is known to happen in low to medium stacking fault energy (SFE) materials whereas CDRX is mostly found in high SFE metals \cite{Sakai2014}. This is explained by the fact that low SFE materials exhibit much less dynamic recovery (DRV) and so are less prompt to formation of low angle grain boundaries (GB that have a disorientation lower than a given threshold, often fixed to $15^{\circ}$) or substructures. Regarding processing parameters, it has been found that low strain rates tend to favor CDRX \cite{Rollett2017}.

Until now, most of the research efforts at the mesoscopic scale have been applied to observe, characterize and model DDRX. This appears natural because common study materials such as stainless steels and nickel-based superalloys undergo those mechanisms under usual range of thermomechanical parameters \cite{Kim2013, Wang2020}. Adding this to the fact that modeling DDRX at mesoscale is more evident since nucleation of recrystallized grains can be implemented naturally within common simulation frameworks whereas modeling CDRX requires to be able to describe substructure formation and evolution. Therefore, one can easily understand why the majority of numerical studies focus on DDRX \cite{Hallberg2010, Beltran2015, Maire2017}.

Recently, several articles focusing on the physical mechanisms underlying CDRX have been published \cite{Souza2019, Liu2021, Qiang2021}. They provide an in depth characterization of CDRX for different alloys and bring some new insight upon substructure formation and evolution. Regarding simulation of CDRX, Gourdet \textit{et al.} \cite{Gourdet2003} have published one of the first article on this topic. They proposed a mean-field model, i.e. where the prediction of main characteristics of the microstructure are averaged and which thus does not directly describe the microstructure topology. They applied this model to simulate CDRX of aluminum alloys. Their model relies on a schematic representation of the microstructure as an aggregate of grains and subgrains. Low angle grain boundaries (LAGB) are grouped into classes of same orientation that undergo a progressive increase in disorientation. If sufficient deformation is applied, these LAGB could possibly transform into high angle grain boundaries (HAGB, i.e. GB that have a disorientation higher than $15^{\circ}$) and contribute to form recrystallized grains. This work still inspires many researchers and some recent works illustrate attempts of expanding it by considering additional phenomena \cite{Maizza2018, Li2020, Buzolin2021} or overcoming some known limitations \cite{McQueen2004}. Those studies have shown the capability of mean-field models of CDRX. Nevertheless, they suffer from the intrinsic limits of mean-field approximations which make impossible the consideration of microstructure heterogeneities.

The present article presents a full-field model of CDRX that offers the possibility to assess the influence of fine microstructural features upon CDRX. The approach employed extends the framework that has been used for several years to model DDRX to the simulation of CDRX \cite{Maire2017, Bernacki2011}. It relies on a level-set (LS) formulation to simulate microstructure evolution during heat treatment and hot deformation. It takes advantage of recent developments regarding simulation of anisotropic grain growth \cite{Fausty2018, Murgas2021} to describe precisely substructure kinetic. It is applied to 2D simulations since an extensive comparison to 2D experimental results will be provided in an upcoming study. Two strategies are employed to introduce substructures. In grain growth (GG) simulations, initial fully substructured microstructures are generated and their evolution is modeled only due to their migration. In dynamic and post dynamic simulations, formation and evolution of subgrains under deformation is tackled by implementing mean-field equations into the LS framework. With those modifications, this paper presents the first numerical framework that is able to model DDRX and CDRX with a high degree of consistency between the physical phenomena considered, the numerical strategies and the material parameters.

The article introduces first the numerical framework, including the basic theory underlying the LS method. Then, generation of digital microstructures and physical phenomena considered are detailed. Following sections are devoted to the presentation of the results of simulation of both GG, CDRX and post-dynamic recrystallization (PDRX). Influence of specific microstructure characteristics are pointed out and discussed in regards of available results in the state of the art.

\section{Numerical framework}\label{sec:NumericalFramework}

\subsection{Grain boundary description}\label{subsec:GBdescription}

The model used in this paper describes GB using a LS formulation.
The iso-zero values of level-set functions $\phi$ are used to track microstructure interfaces in space and time. LS functions are initialized according to equation \ref{eq:InitLSfunction}:

\begin{eqnarray}
   	\begin{cases}
      \phi(x,t) = \pm ~ d\left( x, \Gamma(t) \right) , ~ x \in \Omega, \\
      \Gamma(t) = \lbrace x \in \Omega, ~ \phi(x,t) = 0 \rbrace,
    \end{cases}
\label{eq:InitLSfunction}
\end{eqnarray}

where $d$ is the signed euclidean distance function and $\Omega$ the simulation domain. $\phi(x,t)$ is taken positive inside the grain and negative outside. Initially, one LS function was defined per grain. To reduce the number of functions, a graph coloring strategy is applied \cite{Scholtes2015}. We can note $\Phi = \lbrace \phi_i, ~ i = 1, ~ ..., ~ N\rbrace$ the set of all distance functions used to describe all the grains. $N$ is the number of distance functions and is greatly smaller than the number of grains ($N_g$).

The movement of interfaces is described by solving, for each considered distance function,  the transport equation:

\begin{eqnarray}
 \dfrac{\partial \phi}{\partial t} + \overrightarrow{v}\cdot \overrightarrow{\nabla\phi} = 0,
\label{eq:TransportEquation}
\end{eqnarray}

with $\overrightarrow{v}$ being the velocity of interfaces. At mesoscopic scale, $\overrightarrow{v}$ is generally expressed as the sum of a capillarity term ($\overrightarrow{v}_c$) and of a second one induced by stored energy due to plastic deformation ($\overrightarrow{v}_e$). These two terms are defined such as \cite{Rollett2017, Bernacki2011}:

\begin{eqnarray}
\overrightarrow{v}_c = -M \gamma \kappa \overrightarrow{n} \label{eq:CapillarityVelocity},\\
\overrightarrow{v}_e = M \llbracket E \rrbracket \overrightarrow{n},
\end{eqnarray}

with $M$, $\gamma$ and $\kappa$ respectively the mobility, energy and trace of the curvature tensor of GB. $\llbracket E \rrbracket$ is the stored energy gap between adjacent grains. At grain boundary vicinity, it can be expressed such as: $\llbracket E \left( x, t \right) \rrbracket = f(\phi_i(x,t),l) \times (E_j(t)-E_i(t))$, where f is a decreasing function being equal to 1 for $\phi_i = 0$ to 0 for $\phi_i = l$ and $E_i(t)$ and $E_j(t)$ are the stored energy respectively for grains i and j. Finally, $\overrightarrow{n}$ is the outward unitary normal of GB. The detailed procedure of the computation of $\overrightarrow{v}_e$ is presented in ref. \cite{Bernacki2008}. Stored energy is computed using a dislocation density field:

\begin{eqnarray}
E = \tau \rho
\label{eq:StoredEnergy}
\end{eqnarray}

$\tau$ is the unit dislocation line energy and is defined as a material parameter. Its value is computed thank to following equation: $\tau = \dfrac{\mu b^2}{2}$ with $\mu$ the shear modulus and $b$ the Burgers vector \cite{Dieter1976}.
This field is averaged per grain/subgrain. During simulation, the dislocation density of the surface swept by each interface is reset to a low value ($\rho_0$) and the energy is averaged again. Therefore, the stored energy decreases in grains/subgrains that grow whereas it stays constant for the ones shrinking.

$\kappa$ and $\overrightarrow{n}$ can be defined naturally by taking advantages of the possibilities offered by the LS framework. Indeed, considering that LS functions remain distance functions all along the simulation (i.e. that $\lVert \overrightarrow{\nabla \phi} \rVert = 1$), they can be defined as:

\begin{eqnarray}
\overrightarrow{n} = -\overrightarrow{\nabla \phi}, \label{eq:NormalGB}
\end{eqnarray}

\begin{eqnarray}
\kappa = - \Delta \phi. \label{eq:CurvatureGB}
\end{eqnarray}

$M$ and $\gamma$ are material parameters that should be identified carefully. Setting of those parameters as well as other initial microstructure descriptors (such as initial stored energy) is described in details in section \ref{subsec:MaterialParameters}.

Since CDRX involves significant presence of LAGB, being able to predict their evolution is capital. To do so, evolution of properties with misorientation must be described properly. In the simulations presented in this work, $\gamma$ is always considered as heterogeneous (i.e. $\gamma$ being a function of disorientation, $\gamma=\gamma(\theta)$) and $M$ is either considered homogeneous or heterogeneous, depending on cases. This induces the introduction of an additional term inside velocity equation (eq. \ref{eq:CapillarityVelocity}) \cite{Fausty2018, Murgas2021, Fausty2020}, such that:

\begin{eqnarray}
\overrightarrow{v}_c = -M \left( \overrightarrow{\nabla \gamma} \cdot \overrightarrow{n} - \gamma \kappa \right) \overrightarrow{n},
\end{eqnarray}

which, using equations \ref{eq:NormalGB} and \ref{eq:CurvatureGB} becomes:

\begin{eqnarray}
\overrightarrow{v}_c =  -M \left[ \left( \overrightarrow{\nabla \gamma}  \cdot \overrightarrow{\nabla \phi} \right) \overrightarrow{\nabla \phi} -\gamma \Delta \phi \overrightarrow{\nabla \phi} \right].
\end{eqnarray}

Finally, transport equation becomes:

\begin{eqnarray}
\dfrac{\partial \phi}{\partial t} + \overrightarrow{v}_e\cdot \overrightarrow{\nabla\phi} + M \left[ \overrightarrow{\nabla \gamma}  \cdot \overrightarrow{\nabla \phi} - \gamma \Delta \phi \right] = 0.
\label{eq:TransportEquationFinal}
\end{eqnarray}

The weak formulation of previous equation, with $\varphi \in H_0^1(\Omega)$ is defined as:

\begin{eqnarray}
\begin{split}
\int_{\Omega}  \dfrac{\partial \phi}{\partial t} \varphi d\Omega + \int_{\Omega}  M \gamma \overrightarrow{\nabla \varphi} \cdot \overrightarrow{\nabla \phi} d\Omega  - \int_{\Omega} \overrightarrow{v}_e \cdot \overrightarrow{\nabla \phi} \varphi d\Omega \\
+ 2 \int_{\Omega} M \overrightarrow{\nabla \gamma} \cdot  \overrightarrow{\nabla \phi} \varphi d\Omega + \int_{\Omega} \gamma \overrightarrow{\nabla M} \cdot  \overrightarrow{\nabla \phi} \varphi d\Omega\\
- \int_{\partial \Omega} M \gamma \varphi \overrightarrow{\nabla \phi} \cdot \overrightarrow{n_{\partial \Omega}} d \left(  \partial \Omega \right) = 0.
\end{split}
\end{eqnarray}

$M$ and $\gamma$ are by definition highly discontinuous since they present positive values in elements crossed by GB and are null in all other elements. To avoid numerical issues during computation of $\overrightarrow{\nabla \gamma}$ and $\overrightarrow{\nabla M}$, a Laplace equation is solved to obtain first order differentiable fields with the correct values at GB \cite{Fausty2018}.

This formulation is not the most complete available at that time since some others are able to consider anisotropic grain properties, i.e. that depend also on the GB normal \cite{Florez2022}. However, according to the conclusions of the study made by Murgas \textit{et al.} \cite{Murgas2021}, this heterogeneous formulation remains a good compromise between computation costs and prediction of GB kinetic and morphology.

\subsection{Generation and evolution of microstructures}\label{subsec:SubgrainFormation}

\subsubsection{Generation of a two-levels microstructure}\label{subsubsec:TwoLevelsMicrostructureGeneration}

A first method to obtain a microstructure exhibiting substructures is to employ a two-levels generation method. The generation strategy relies on a Laguerre-Voronoï algorithm \cite{Hitti2012} using an optimized sphere packing strategy \cite{Hitti2013}. This generation algorithm is called twice. First, it is used to generate the grain structure in a standard way. Then, it is called again to generate the subgrain structure. The nuclei of the Voronoï cells located too close to initial GB are removed. Then, LS functions are defined based on those cells.

\begin{figure}
	\centering
 	\includegraphics[width=0.75\textwidth]{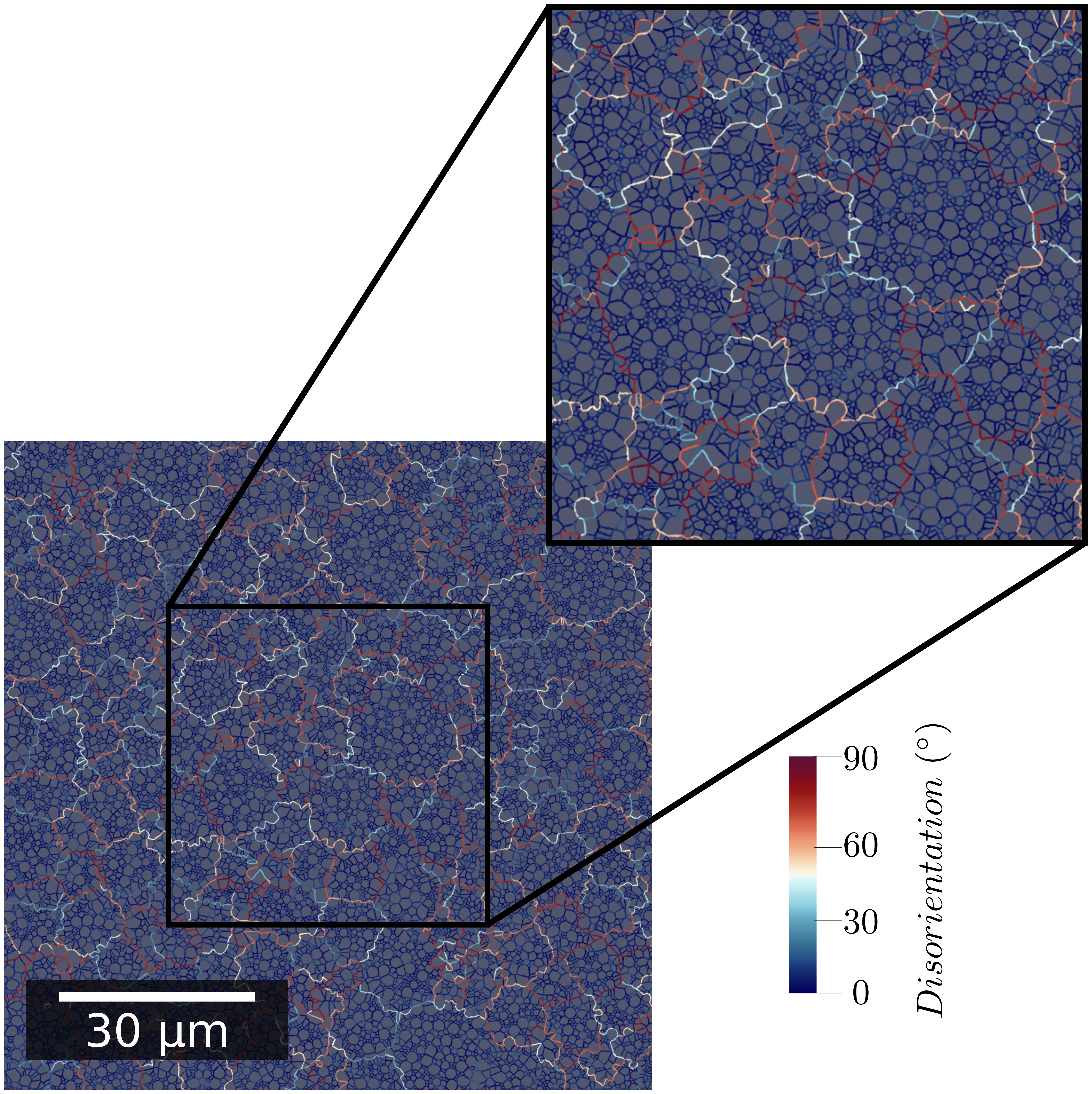}
  \caption{Initial microstructure example with color code corresponding to GB misorientation angle.}\label{fig:InitialMicrostructure}
\end{figure}

The generation algorithm is in charge of respecting the prescribed grain and subgrain size distributions. Grain orientations are affected in order to respect a given distribution taken from experimental data obtained on a zircaloy-4 sample. Subgrain orientation is computed by applying a specific misorientation to the parent grain orientation, similarly to what has been achieved in ref. \cite{Despres2020}. This misorientation distribution is also taken from experimental data. Finally, an image of an initial microstructure is shown in figure \ref{fig:InitialMicrostructure}. The zone at the center that is enlarged is kept the same for figure \ref{fig:EvolutionGBnetworkGG} to improve the readability.

It is worth noting that this method allows to generate various initial configurations and to study in details the influence of topology. As illustrated in section \ref{subsec:ModelGG}, this enables to enrich the rare existing discussions about the full-field modeling of the evolution of subgrain structures during grain growth \cite{Despres2020, Holm2003, Suwa2008}.

\subsubsection{Progressive formation of subgrains}\label{subsubsec:ProgressiveSubgrainFormation}

As described in introduction (section \ref{sec:Intro}), under deformation, LAGB network does not evolve only by capillarity growth. Dislocations could rearrange themselves to form new LAGB or accumulate into preexisting LAGB. This last phenomenon is responsible for a progressive increase of LAGB misorientation \cite{Rollett2017, Huang2016}. To be able to consider these mechanisms, laws introduced by the Gourdet-Montheillet model are implemented \cite{Gourdet2003}.

Each grain has an average dislocation density $\rho$ which evolves according to the Yoshie-Laasraoui-Jonas equation \cite{Laasraoui1991}:

\begin{eqnarray}
d\rho = \left( K_1 - K_2 \rho \right) d\varepsilon,
\label{eq:YLJ}
\end{eqnarray}

where $K_1$ and $K_2$ are two material constants respectively describing the strain hardening and the recovery. To be able to reproduce to some extent heterogeneous grain deformation, $K_1$ varies from grain to grain according to a distribution defined using experimental data (see section \ref{subsec:MaterialParameters}). This strategy is preferred over the coupling with a crystal plasticity model since it would increase unreasonably computational costs.
Several mechanisms of dislocation density evolution are taken into account into the current model:
\begin{itemize}
    \item rearrangement into LAGB that bound new subgrains. Subgrain formation is described through following equation \cite{Gourdet2003}:
    \begin{eqnarray}
    dS^+ = \dfrac{\alpha b K_2 \rho d\varepsilon}{\eta \theta_0},
    \label{eq:SurfaceNewSubgrains}
    \end{eqnarray}
    where $dS^+$ is the surface of LAGB created. $\alpha = 1- \exp{\left( \dfrac{D}{D_0} \right) ^m}$ is a coefficient describing the fraction of dislocations recovered to form new subgrains. $D$ is the grain diameter, $D_0$ is a grain reference diameter and $m$ is a fixed coefficient. $\eta$ is the number of sets of dislocations and $\theta_0$ the disorientation of newly formed subgrains.
    \item Stacking into preexisting LAGB which is modeled according to the following equation \cite{Gourdet2003}:
    \begin{eqnarray}
    d\theta = \dfrac{b}{2 \eta} \left(1-\alpha\right) D K_2 \rho d\varepsilon.
    \label{eq:ProgressiveMisorientation}
    \end{eqnarray}
    \item Absorption during HAGB migration. This is naturally captured by affecting to the areas swept by moving boundaries a low dislocation density as described earlier.
\end{itemize}

At each time step, dislocation density of each grain is updated using equation \ref{eq:YLJ} which impacts the computation of the velocity term related to stored energy differences. Then, the length of subgrain interfaces formed into each grain is computed using equation \ref{eq:SurfaceNewSubgrains}. Depending on the simulation case (see section \ref{subsec:CDRX}), subgrains are added grain by grain based on the value of this grain property or globally, after having summed the length of subgrain interfaces for all grains. Subgrain orientation is initialized by applying a small misorientation to the parent grain orientation (similarly to what is described in section \ref{subsubsec:TwoLevelsMicrostructureGeneration}). The misorientation angle is selected to respect a distribution measured experimentally whereas the misorientation axis satisfies a uniform distribution. The misorientation axis attributed to a subgrain at its formation is kept constant. Then, during next increments, the misorientation increase described by equation \ref{eq:ProgressiveMisorientation} is realized by rotating of $d\theta$ around this axis.
In addition, to evaluate the influence of subgrain formation localization, a distance from pre-existing boundaries criterion is tested. Therefore, for the cases where this option is enabled, subgrain centers cannot be set closer to pre-existing interfaces than $d^f_{\textsc{sg}}$. This will be discussed in more details in section \ref{subsec:CDRX}.

Finally, at each time step, isotropic remeshing is performed to keep a fine mesh around GB and precise description of interfaces \cite{Bernacki2009, Roux2013}.
All those operations are presented in an algorithmic diagram in figure \ref{fig:CDRXalgo}.

\subsection{Material parameters}\label{subsec:MaterialParameters}

GB energy is defined according to Read-Schockley equation \cite{Read1950}:

\begin{eqnarray}
\gamma  (\theta)  =
   	\begin{cases}
     \gamma_{max} \left( \dfrac{\theta}{\theta_{max}}\right)\left(1- \ln{\dfrac{\theta}{\theta_{max}}}\right) , ~ \theta < \theta_{max}, \\
      \gamma_{max}, ~ \theta \geq \theta_{max},
    \end{cases}
\label{eq:GammaRS}
\end{eqnarray}

where $\theta_{max}$ is the limit between LAGB and HAGB, taken equal to $15 ^{\circ}$.

If not explicitly written, GB mobility, $M$, is considered isotropic and does not vary with disorientation (i.e. $M = M_{max}$). HAGB mobility ($M_{max}$) is identified using experimental data. Its dependence to temperature is supposed to follow an Arrhenius law. In some given simulation cases and to be able to discuss and compare the results to the ones described by Suwa \textit{et al.} \cite{Suwa2008}, mobility is described by the following relation \cite{Huang2000}:

\begin{eqnarray}
M(\theta) =
   	\begin{cases}
    M_{max} \left( 1- \exp{ \left[-5 \left( \dfrac{\theta}{\theta_{max}} \right)^4 \right]} \right), \\
    M_{max} , ~ \theta \geq \theta_{max}.
    \end{cases}
\label{eq:Mobility}
\end{eqnarray}

\section{Results and discussion}\label{sec:Results}

\subsection{Modeling of GG of a fully substructured microstructure}\label{subsec:ModelGG}

\subsubsection{Influence of microstructure topology}

\begin{figure}
    \begin{subfigure}{0.5\textwidth}
        \centering
        \includegraphics[width=0.95\linewidth]{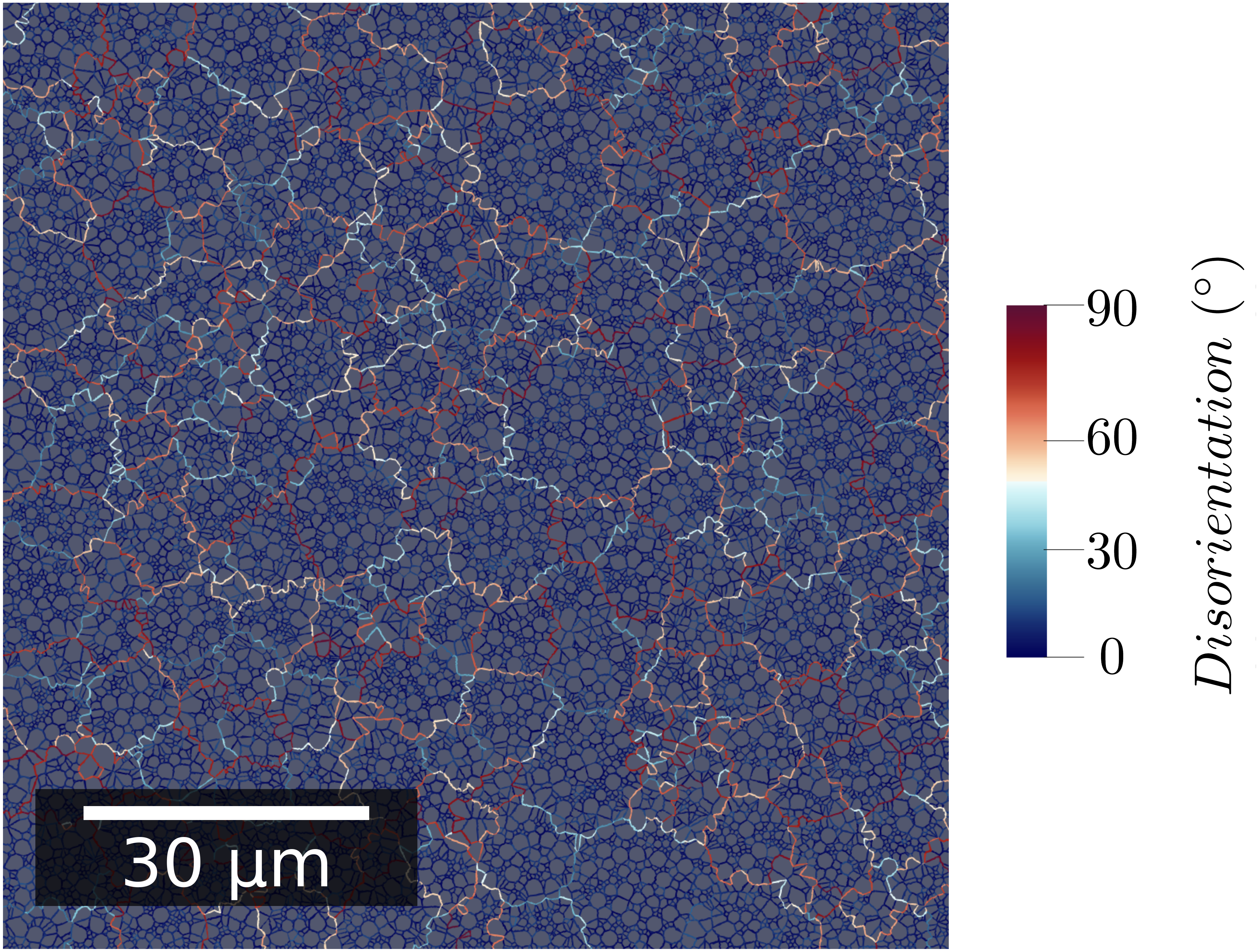}
        \caption{\label{fig:InitialMicrostructureStandardTopology} Standard grain/subgrain topology.}
    \end{subfigure}
    \begin{subfigure}{0.5\textwidth}
        \centering
        \includegraphics[width=0.95\linewidth]{./Figures/Fig2b.pdf}
        \caption{\label{fig:InitialMicrostructureNoTopology} No grain/subgrain topology.}
    \end{subfigure}
\caption{Initial GB network with color code corresponding to GB disorientation.}
\label{fig:InitialMicrostructuresTopology}
\end{figure}

To evaluate the influence of microstructure topology, two different initial microstructures are considered:
\begin{itemize}
    \item with subgrains located inside grains using the generation method described in section \ref{subsubsec:TwoLevelsMicrostructureGeneration} (figure \ref{fig:InitialMicrostructureStandardTopology}),
    \item with LAGB and HAGB evenly distributed throughout the whole representative volume element (RVE) (see figure \ref{fig:InitialMicrostructureNoTopology}).
\end{itemize}

The first microstructure topology is named in the following discussions as \textit{Standard grain/subgrain topology}. The second is named \textit{No grain/subgrain topology}. The two initial microstructures are presented in the following figure \ref{fig:InitialMicrostructuresTopology}

For all simulation cases presented in this section, GB energy is taken heterogeneous and GB mobility is considered isotropic, stored energy is considered negligible (i.e. $v_e = 0$). The RVE area is equal to $0.01 ~ mm^2$.

\begin{figure}
	\centering
 	\includegraphics[width=0.65\textwidth]{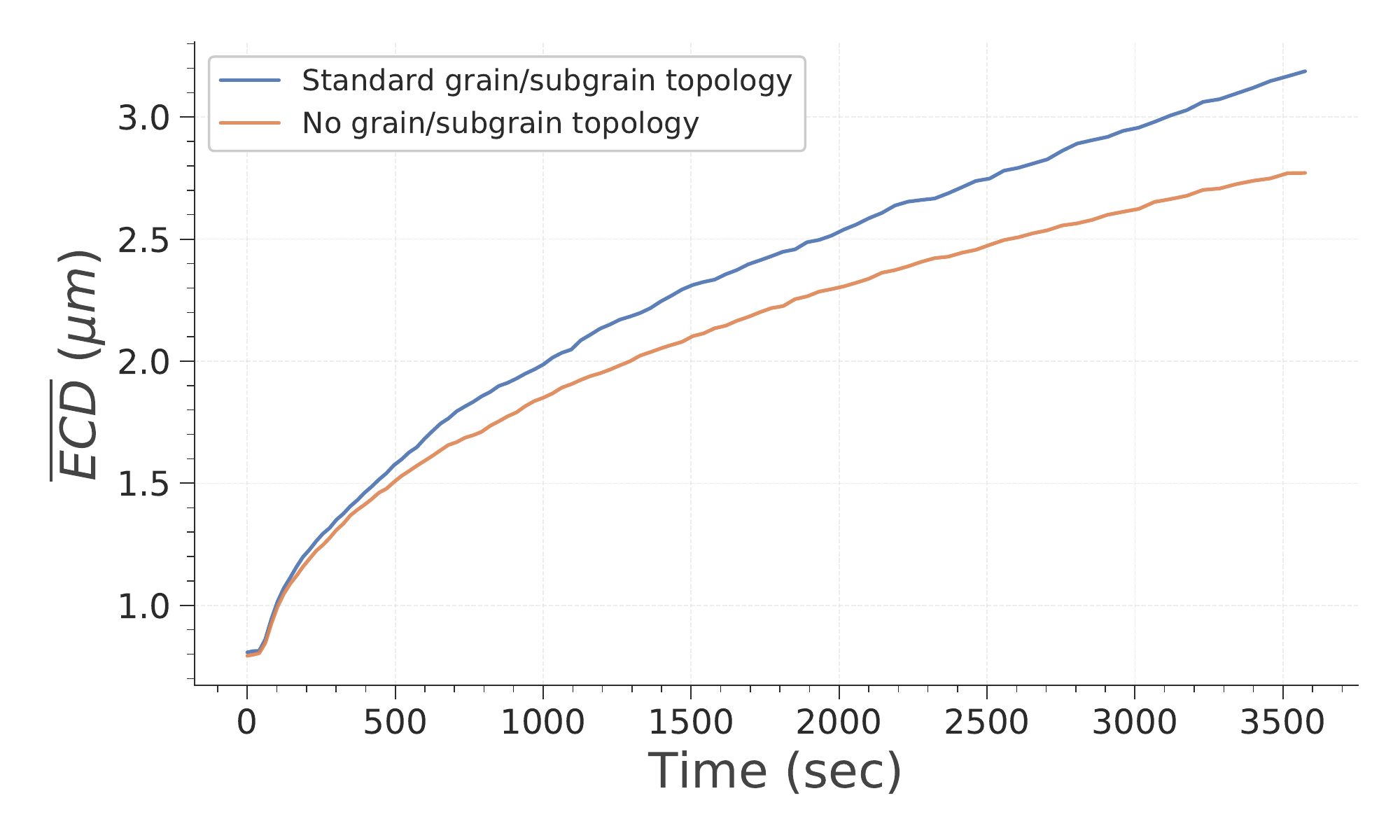}
  \caption{Evolution of subgrain $\overline{ECD}$ as a function of time.}\label{fig:GG_ECD_topology}
\end{figure}

\begin{figure}
    \begin{subfigure}{0.5\textwidth}
        \centering
        \includegraphics[width=0.95\linewidth]{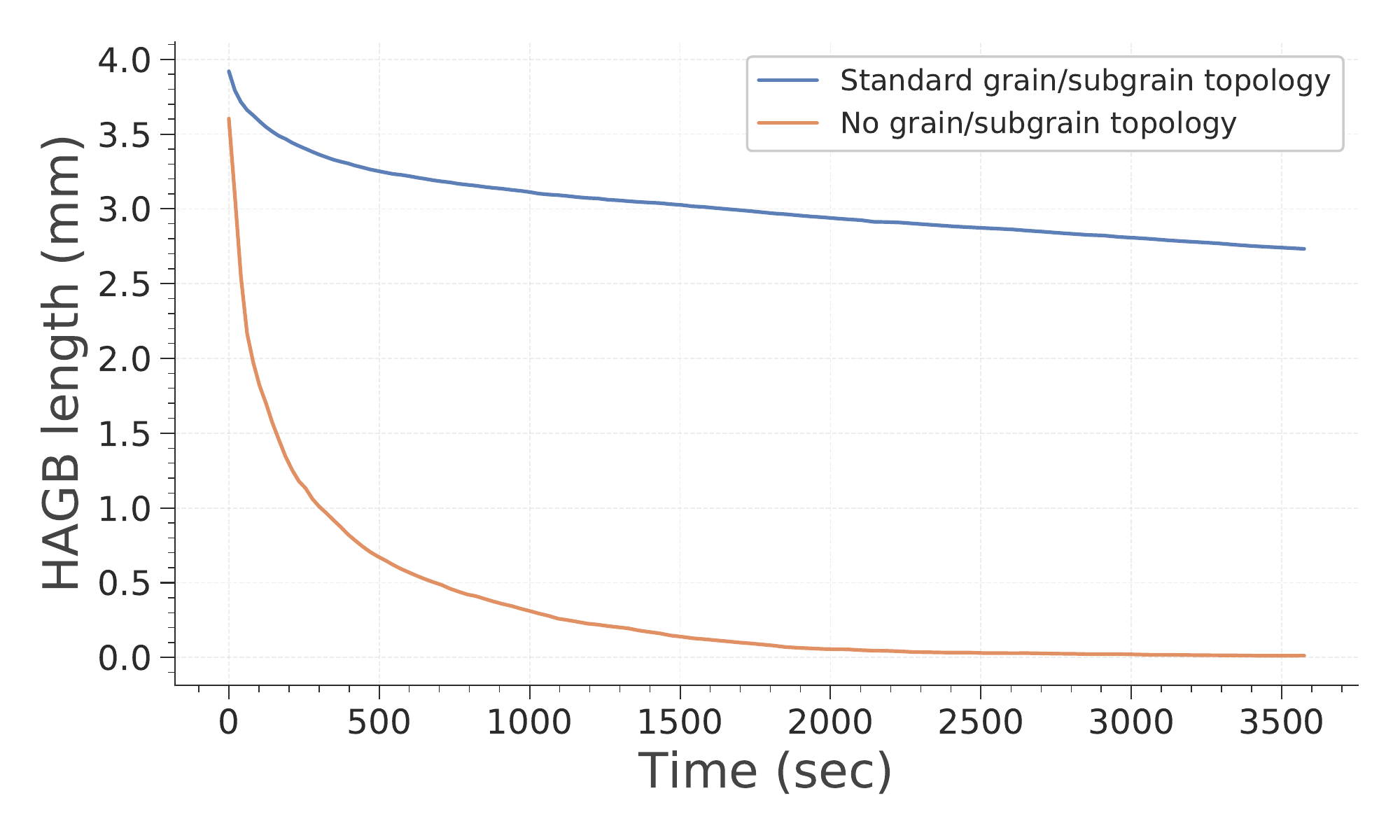}
        \caption{\label{fig:HAGBlengthratioTopology} Total HAGB length.}
    \end{subfigure}
    \begin{subfigure}{0.5\textwidth}
        \centering
        \includegraphics[width=0.95\linewidth]{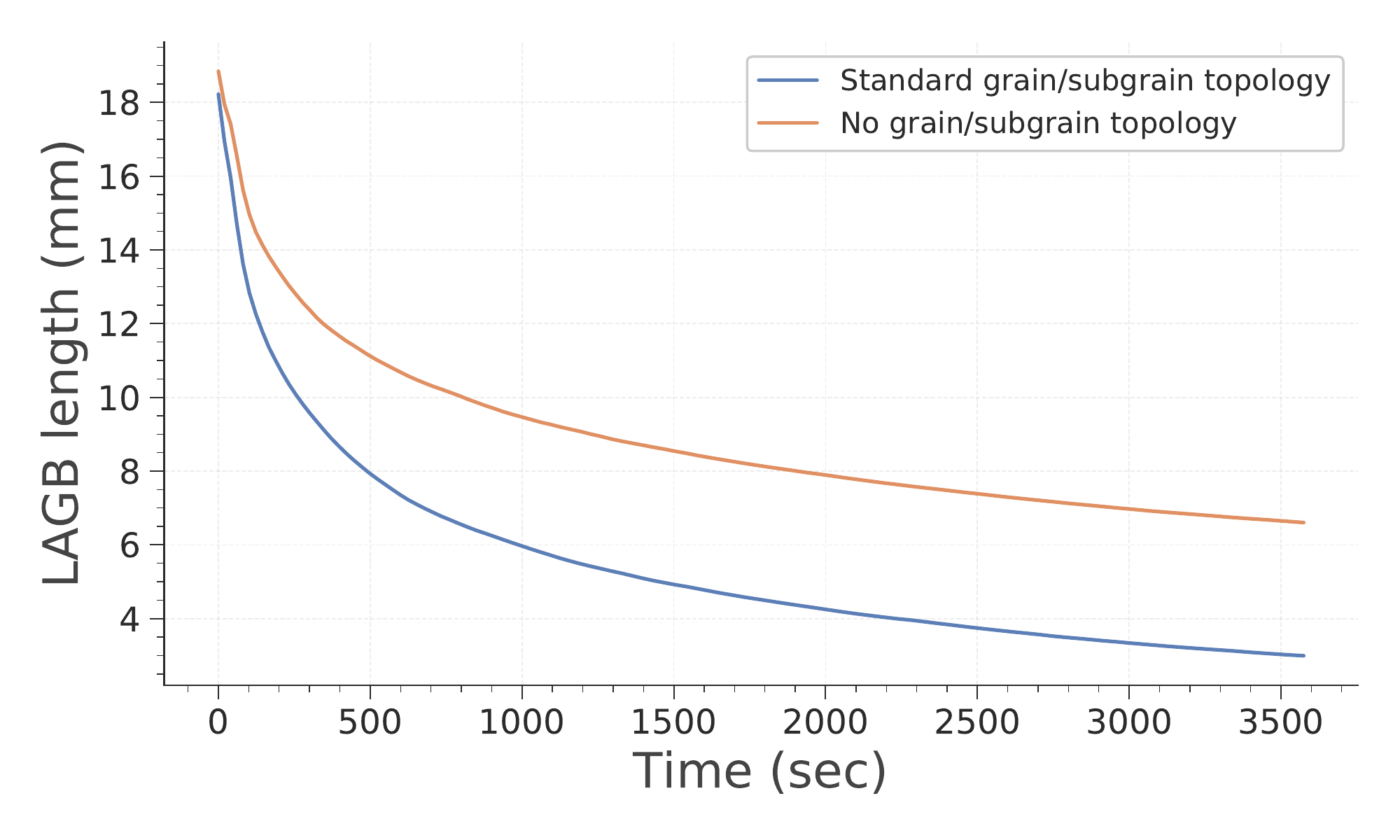}
        \caption{\label{fig:LAGBlengthTopology} Total LAGB length.}
    \end{subfigure}\vspace{0.3cm}
    \begin{subfigure}{0.5\textwidth}
        \centering
        \includegraphics[width=0.95\linewidth]{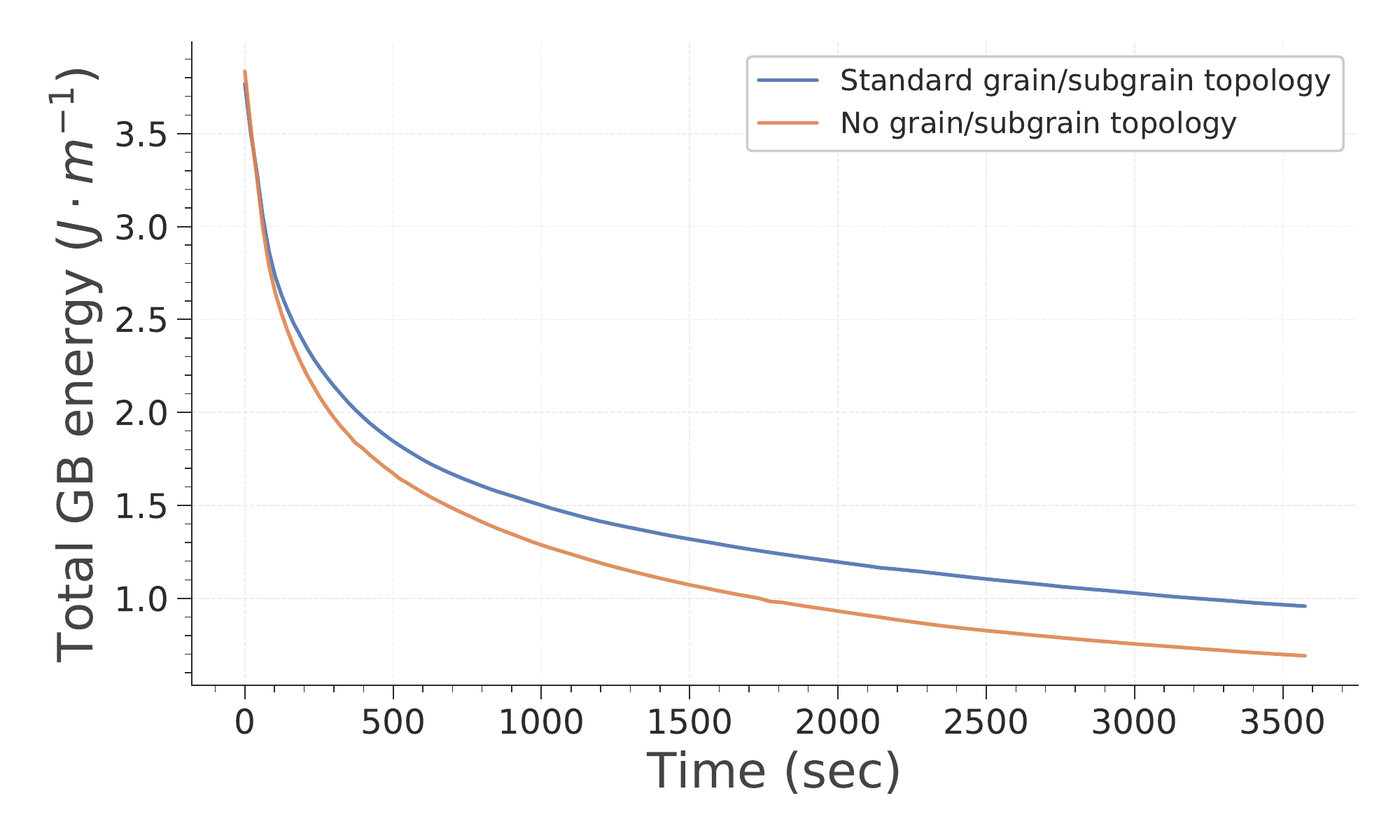}
        \caption{\label{fig:GBenergy} Total GB energy.}
    \end{subfigure}
\caption{Evolution of GB length and energy with time.}
\label{fig:GBlengthsTopology}
\end{figure}

Figures \ref{fig:GG_ECD_topology} and \ref{fig:GBlengthsTopology} point out how spatial correlation of HAGB and LAGB impacts microstructure evolutions. First, it is interesting to note that the absence of spatial correlation decreases grain growth kinetic. This can be explained partly by the fact that grains (i.e. bounded by HAGB) disappear faster when no correlation is assumed (as illustrated in figure \ref{fig:HAGBlengthratioTopology}). Therefore, this leads to lower the global grain growth kinetic since the global system energy is initially lowered much faster. It is interesting to note that this behavior is different from the one expected by Desprès \textit{et al.} \cite{Despres2020}. Two reasons can lead to these differences. First, Desprès \textit{et al.} \cite{Despres2020} consider subgrains partly bounded by HAGB. Second, initial subgrain distribution is different from our study case, which could also impact the microstructure evolution.

Finally, when considering the influence of spatial distribution of orientations and misorientations, one limitation of this work should be pointed out. Indeed, using the generation method presented earlier, it is not possible to generate subgrains inside a grain presenting gradient of misorientations, a feature which is commonly observed in experiments and which could also impacts the microstructure evolutions. This could for instance lead to speed up the formation of HAGB by the migration of LAGB that could end up in the contact between two subgrains with a disorientation greater than $15 ^\circ$. Nevertheless, this limitation could be overcome by directly immersing experimental data.

\subsubsection{Influence of subgrain parameters}

To assess the influence of subgrain parameters, two different initial microstructures are considered:
\begin{itemize}
    \item with grain and subgrain size distributions taken from experimental data,
    \item with grain distribution taken from experimental data and a unique subgrain size.
\end{itemize}
For each of these initial microstructures, two cases are considered in which some material parameters differ:
\begin{itemize}
    \item GB energy is taken heterogeneous and GB mobility is considered constant,
    \item GB energy and mobility are both considered heterogeneous.
\end{itemize}
In these cases, the standard grain/subgrain topology is assumed and stored energy is neglected  (i.e. $v_e = 0$).

\begin{figure}
    \begin{subfigure}[t]{0.5\textwidth}
        \centering
        \includegraphics[width=0.95\linewidth]{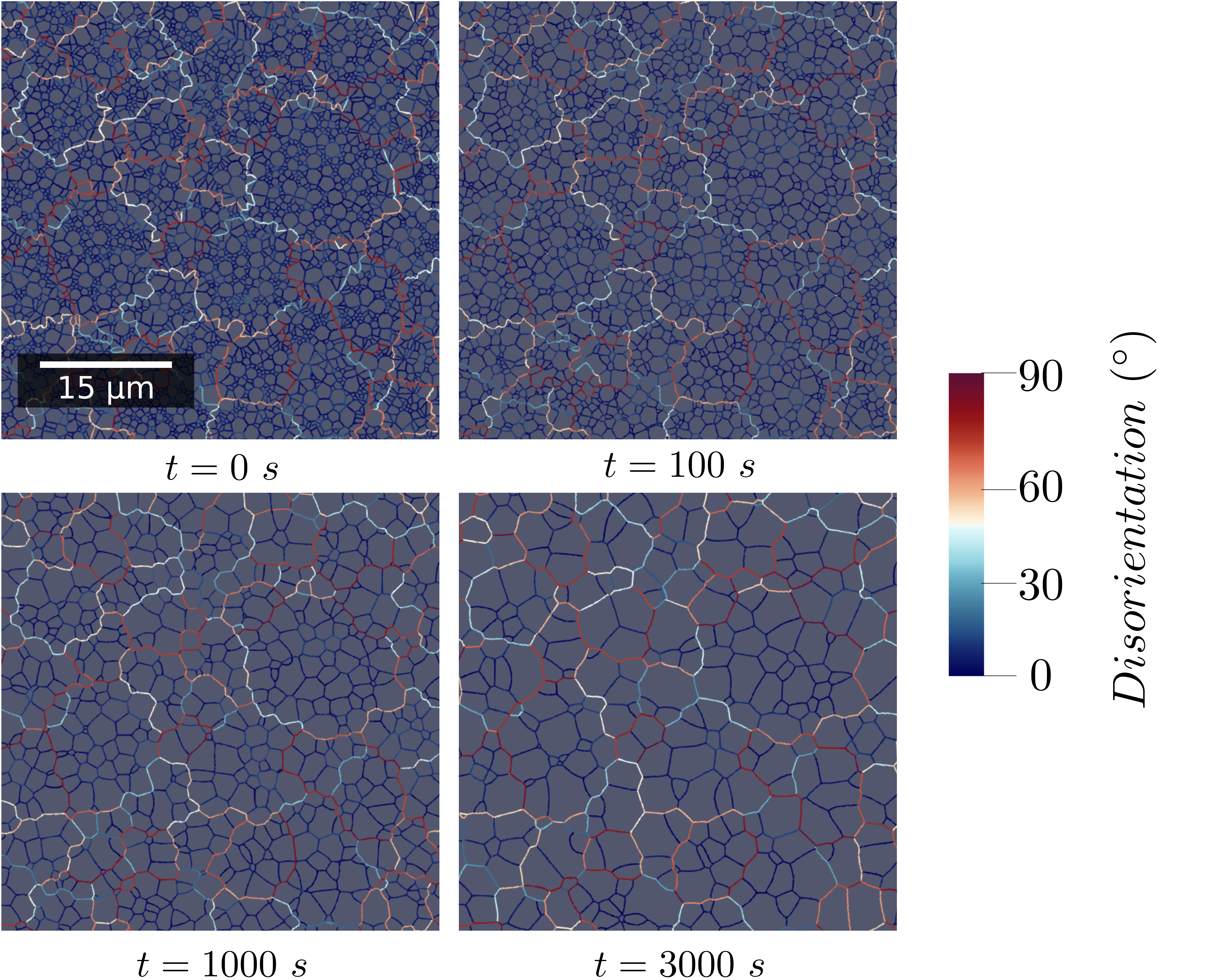}
        \caption{\label{fig:GG_HetGamma} Experimental subgrain size distribution - $\gamma_h$.}
    \end{subfigure}
    \begin{subfigure}[t]{0.5\textwidth}
        \centering
        \includegraphics[width=0.95\linewidth]{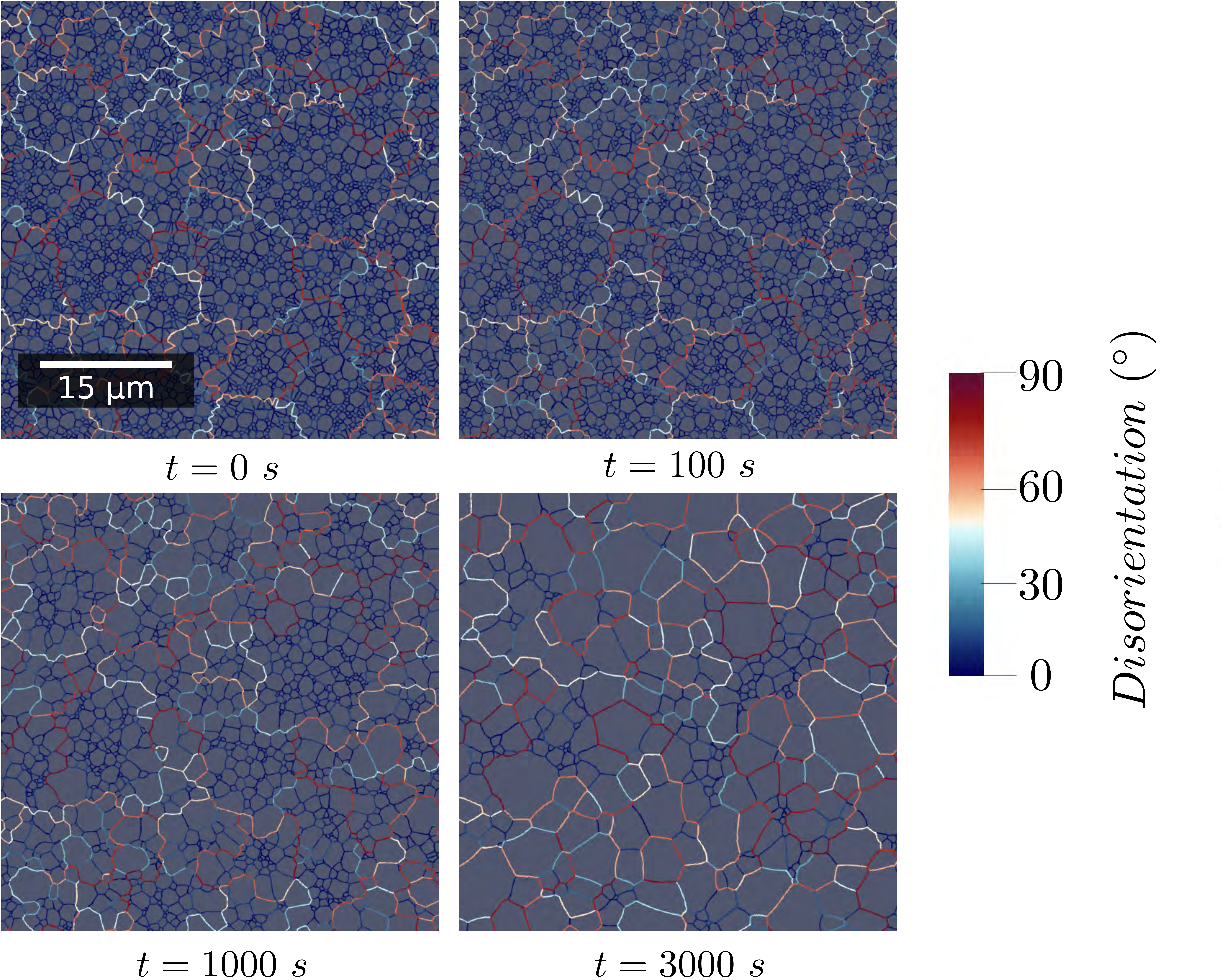}
        \caption{\label{fig:GG_HetGammaHetMob} Experimental subgrain size distribution - $\gamma_h, ~ M_h$.}
    \end{subfigure}\vspace{0.3cm}
    \begin{subfigure}[t]{0.5\textwidth}
        \centering
        \includegraphics[width=0.95\linewidth]{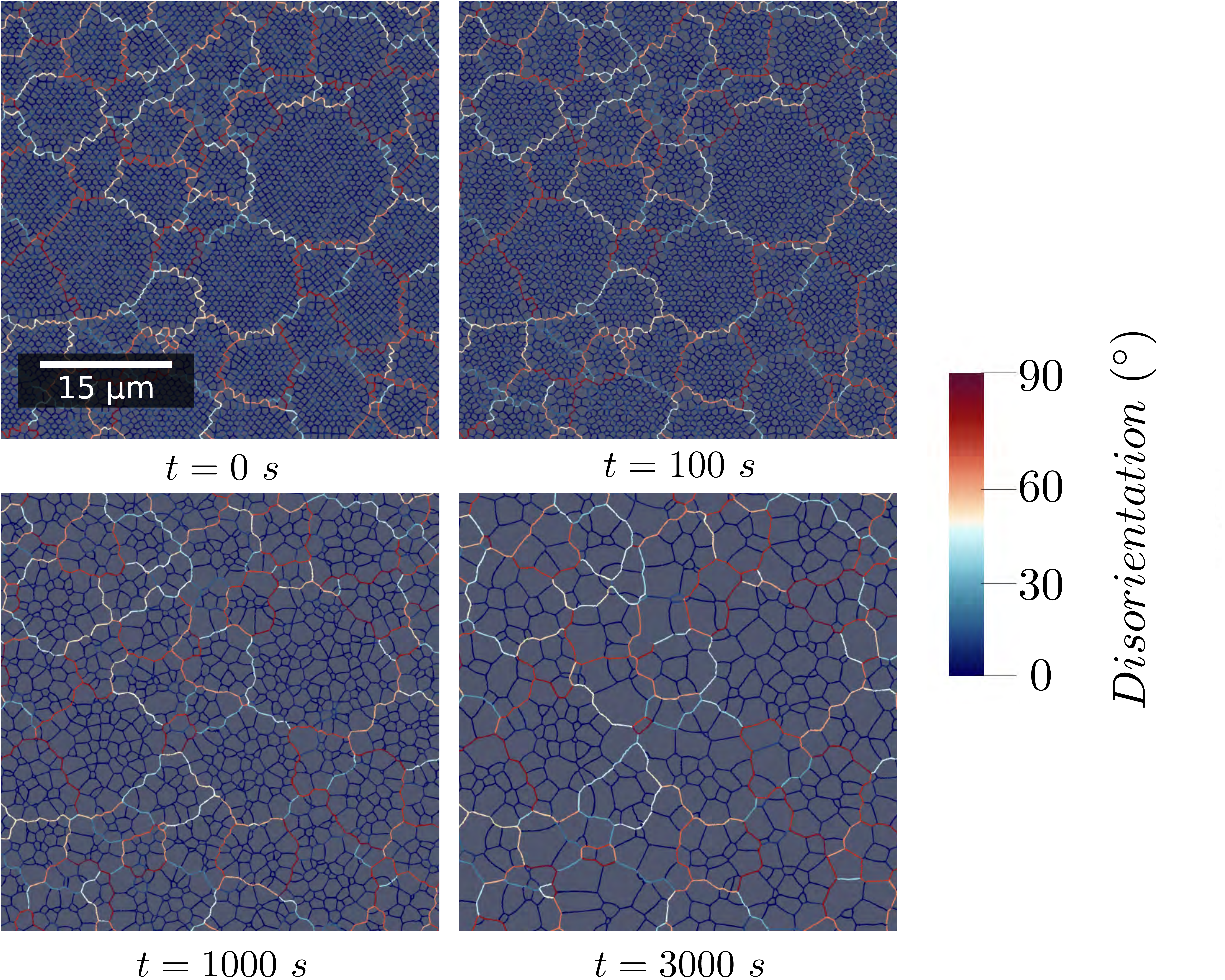}
        \caption{\label{fig:GG_HetGammaUniformSubgrainSize} Uniform subgrain size distribution - $\gamma_h$.}
    \end{subfigure}
    \begin{subfigure}[t]{0.5\textwidth}
        \centering
        \includegraphics[width=0.95\linewidth]{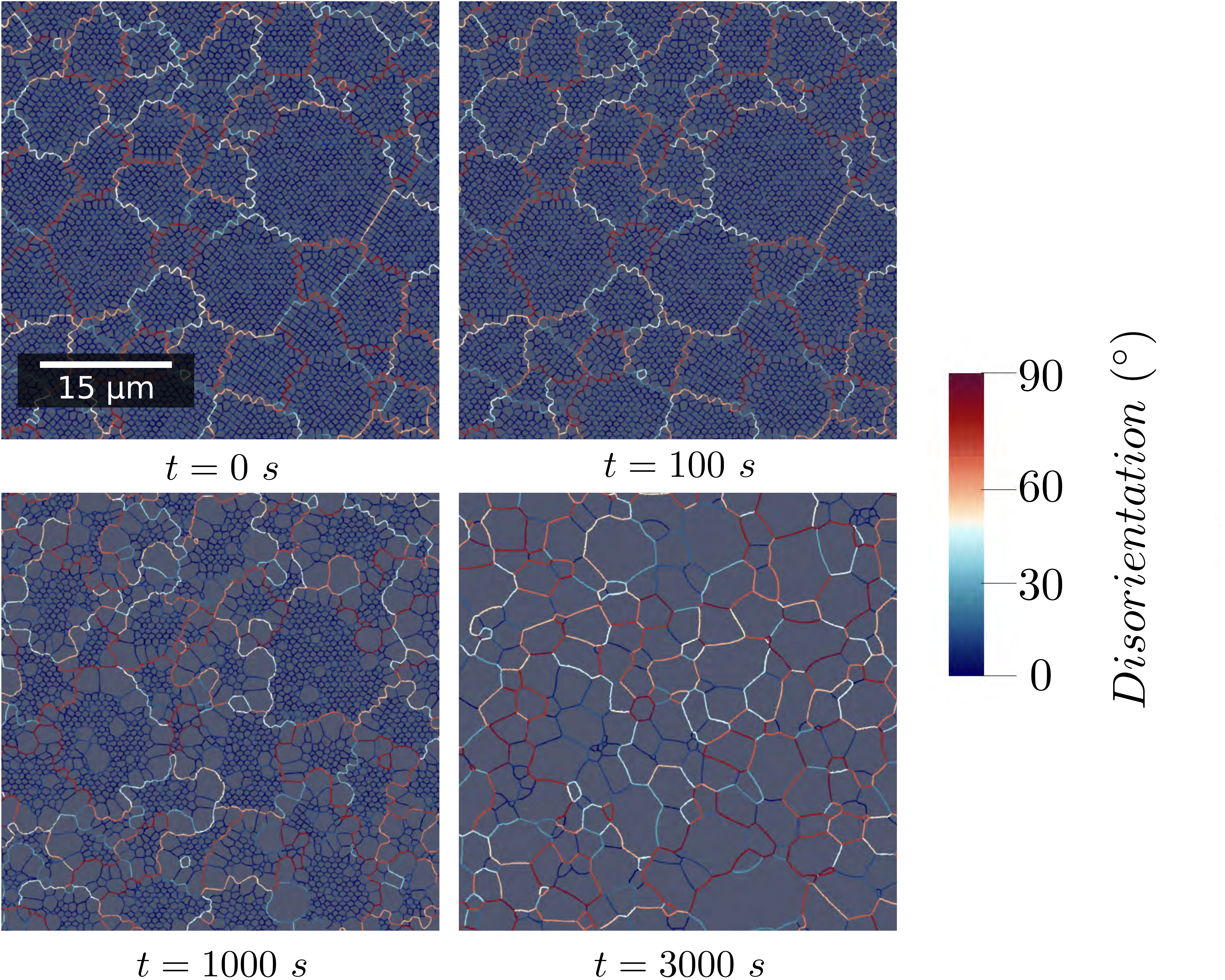}
        \caption{\label{fig:GG_HetGammaHetMobUniformSubgrainSize} Uniform subgrain size distribution - $\gamma_h, ~ M_h$.}
    \end{subfigure}
\caption{Evolution of the GB network with time for the four test cases, in a zone at the center of the RVE.}
\label{fig:EvolutionGBnetworkGG}
\end{figure}

Figure \ref{fig:EvolutionGBnetworkGG} presents for each simulation case the GB network in a part of the RVE. Evolution of the whole RVE are available in appendix \ref{app:EvolutionGBnetworkFullRVE}. This illustrates how microstructure evolves in a noticeable manner in case $(d)$.

\begin{figure}
	\centering
 	\includegraphics[width=0.65\textwidth]{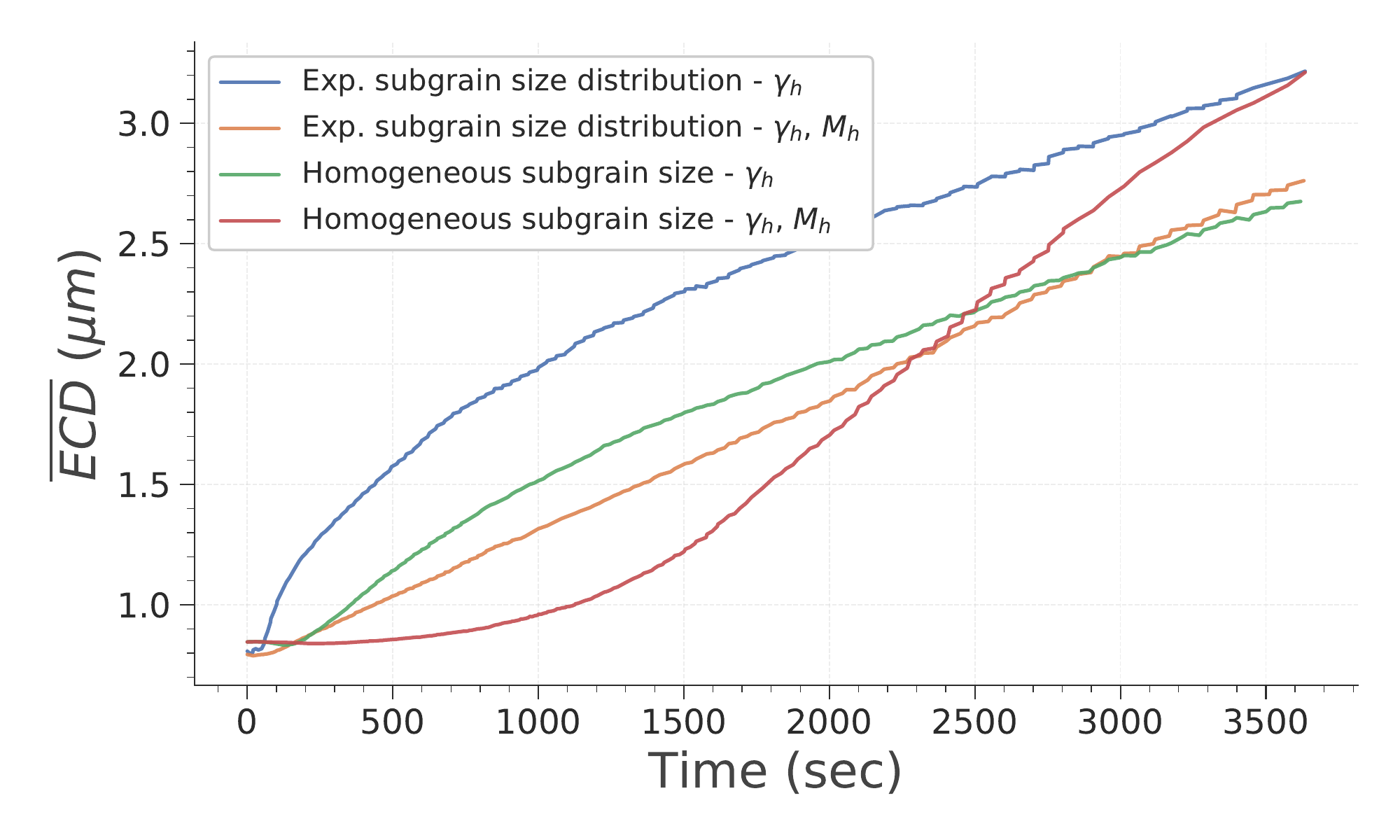}
  \caption{Evolution of subgrain average ECD as a function of time.}\label{fig:GG_ECD}
\end{figure}

\begin{figure}
    \begin{subfigure}{0.49\textwidth}
        \centering
        \includegraphics[width=0.95\linewidth]{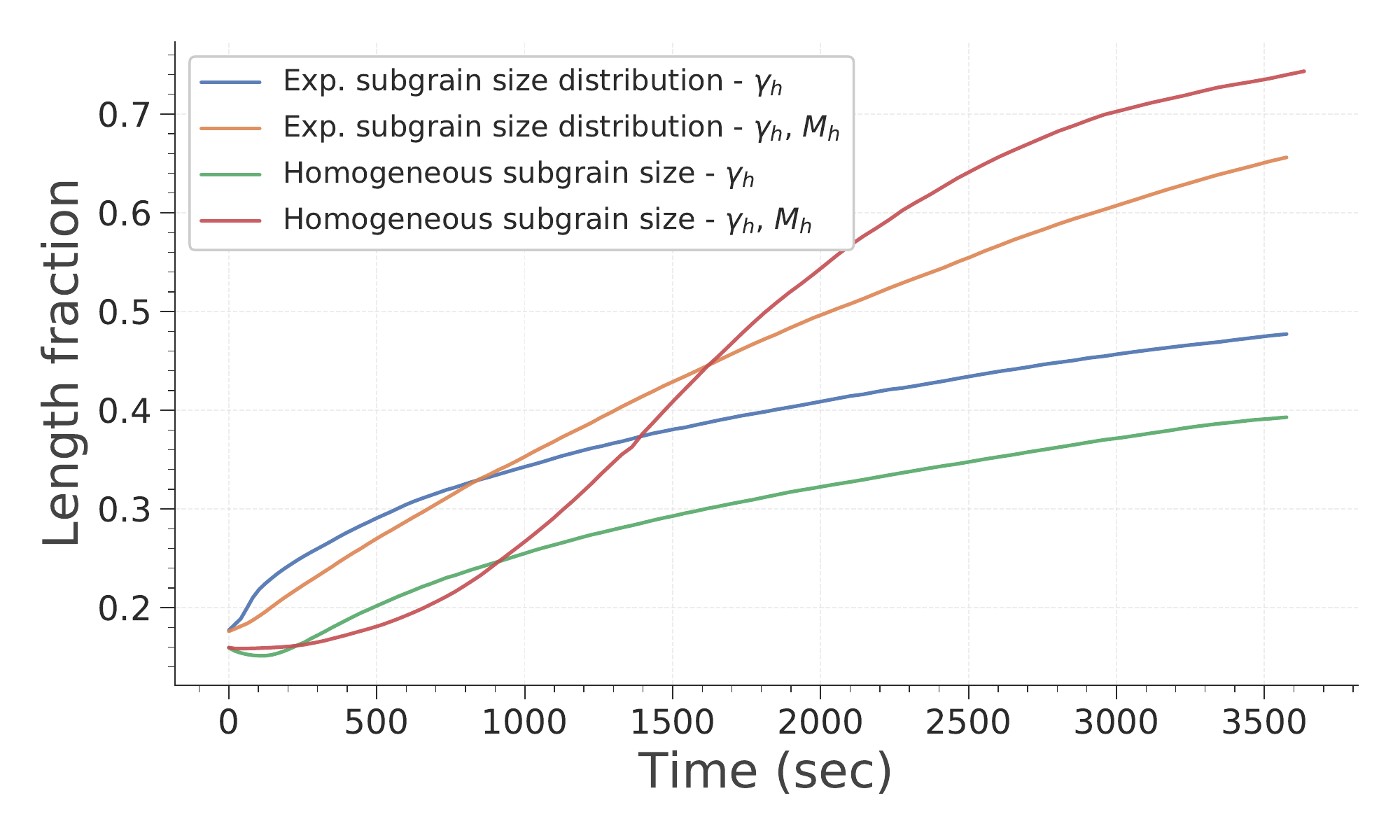}
        \caption{\label{fig:HAGBlengthratio} HAGB length ratio.}
    \end{subfigure}
    \begin{subfigure}{0.49\textwidth}
        \centering
        \includegraphics[width=0.95\linewidth]{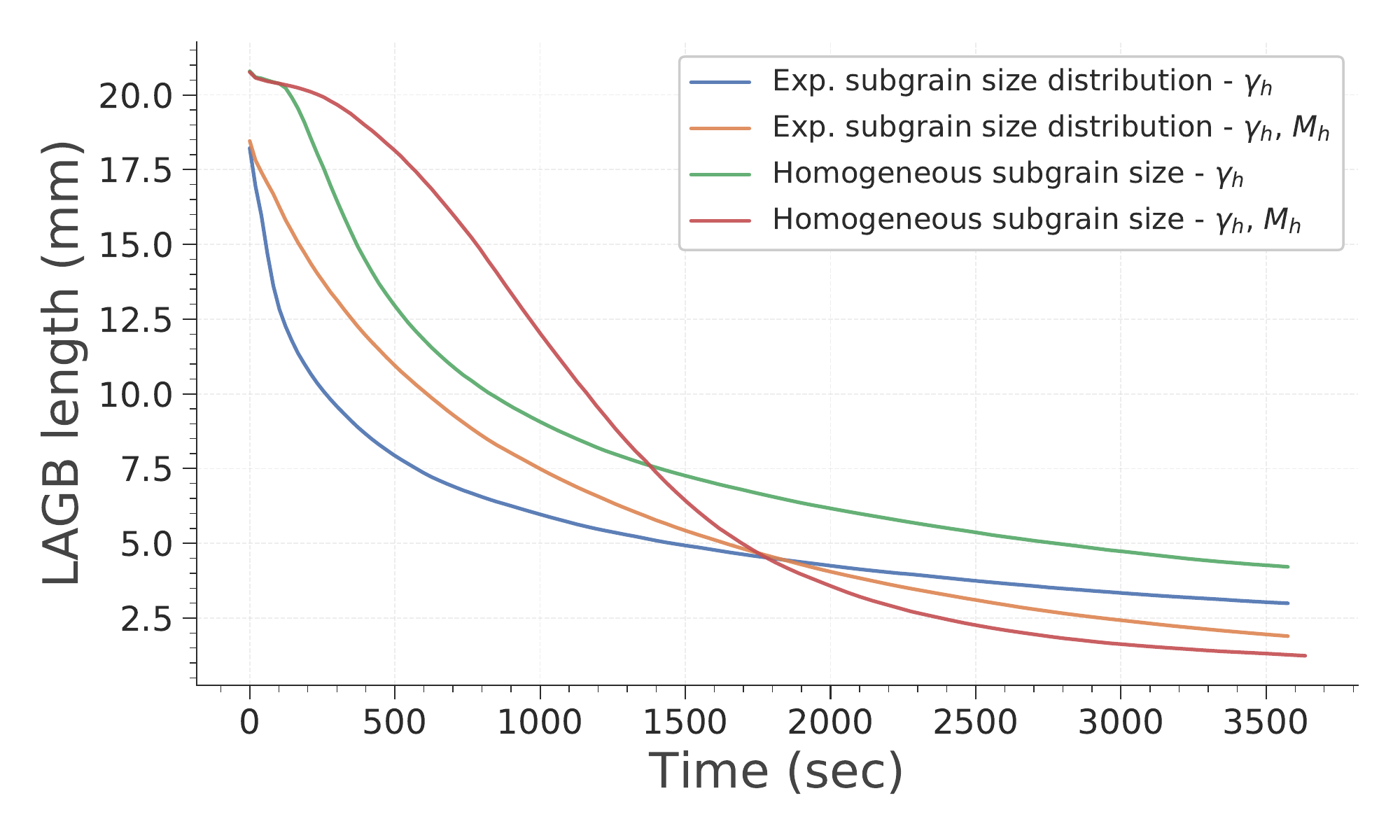}
        \caption{\label{fig:LAGBlength} Total LAGB length.}
    \end{subfigure}\vspace{0.3cm}
    \begin{subfigure}{0.49\textwidth}
        \centering
        \includegraphics[width=0.95\linewidth]{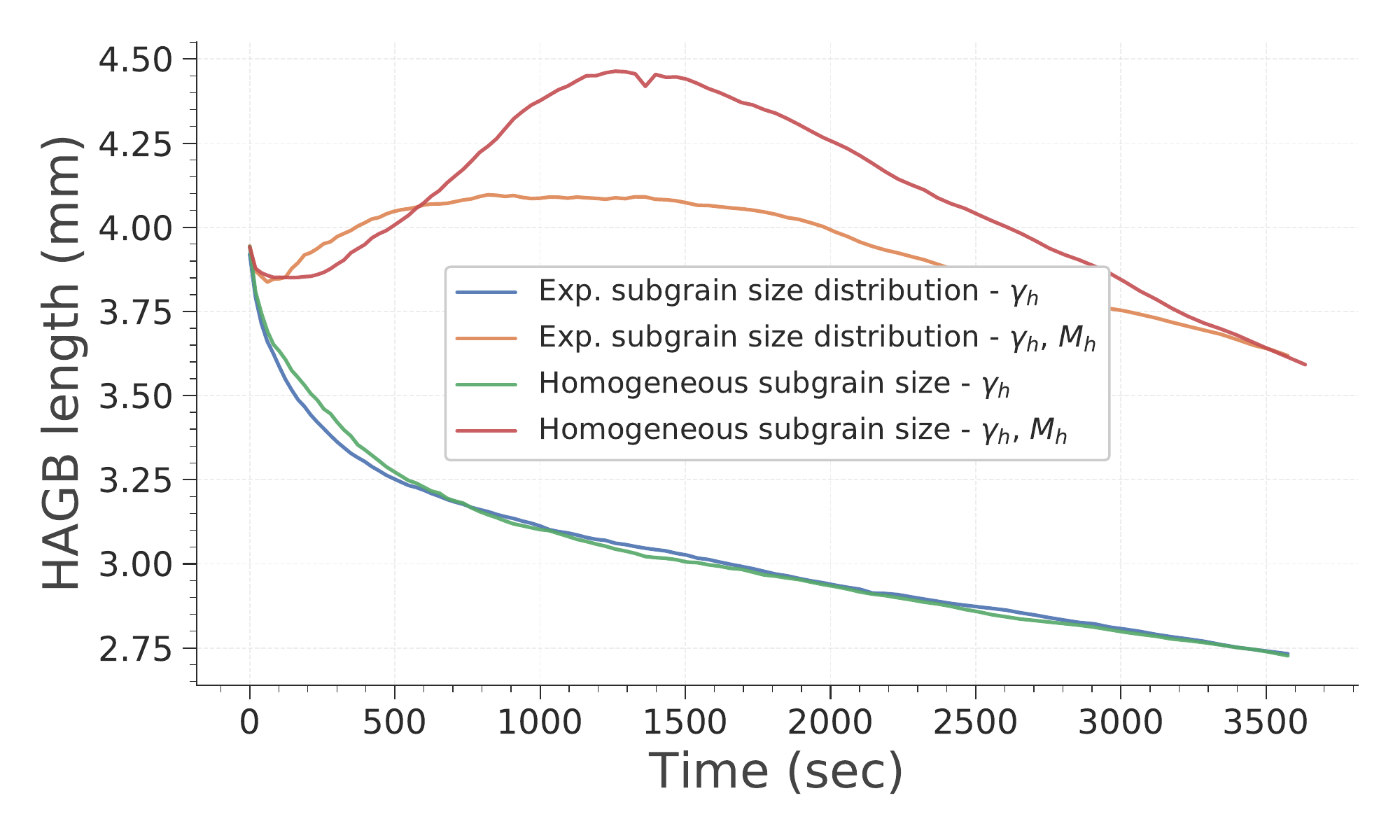}
        \caption{\label{fig:HAGBlength} Total HAGB length.}
    \end{subfigure}
\caption{Evolution of GB length with time.}
\label{fig:GBlengths}
\end{figure}

\begin{figure}
    \begin{subfigure}{0.49\textwidth}
        \centering
        \includegraphics[width=0.95\linewidth]{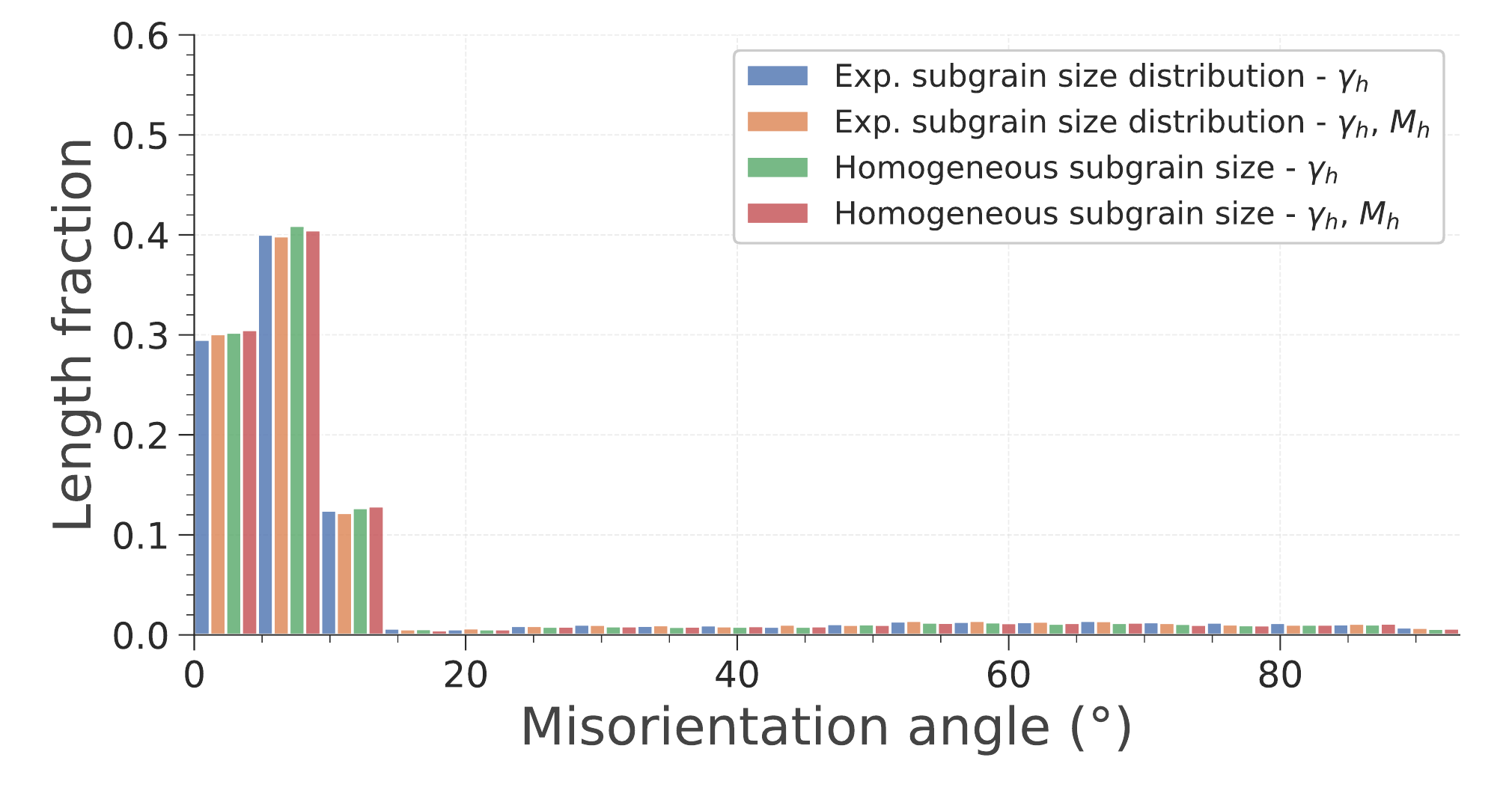}
        \caption{\label{fig:InitialDisorientationDistributions} Initial state.}
    \end{subfigure}
    \begin{subfigure}{0.49\textwidth}
        \centering
        \includegraphics[width=0.95\linewidth]{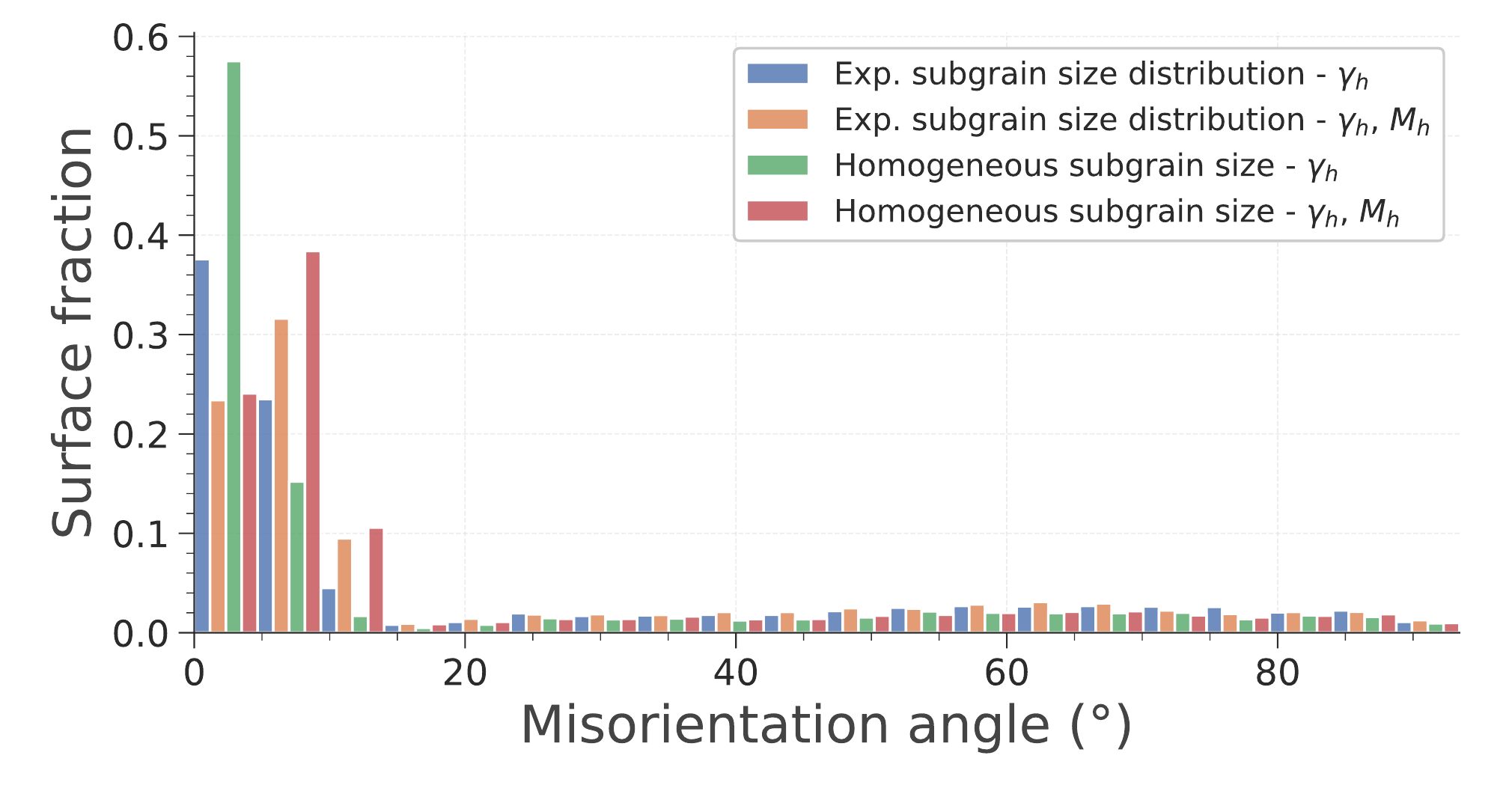}
        \caption{\label{fig:IntermediateDisorientationDistributions} After $1000 ~ s$.}
    \end{subfigure}\vspace{0.3cm}
    \begin{subfigure}{0.49\textwidth}
        \centering
        \includegraphics[width=0.95\linewidth]{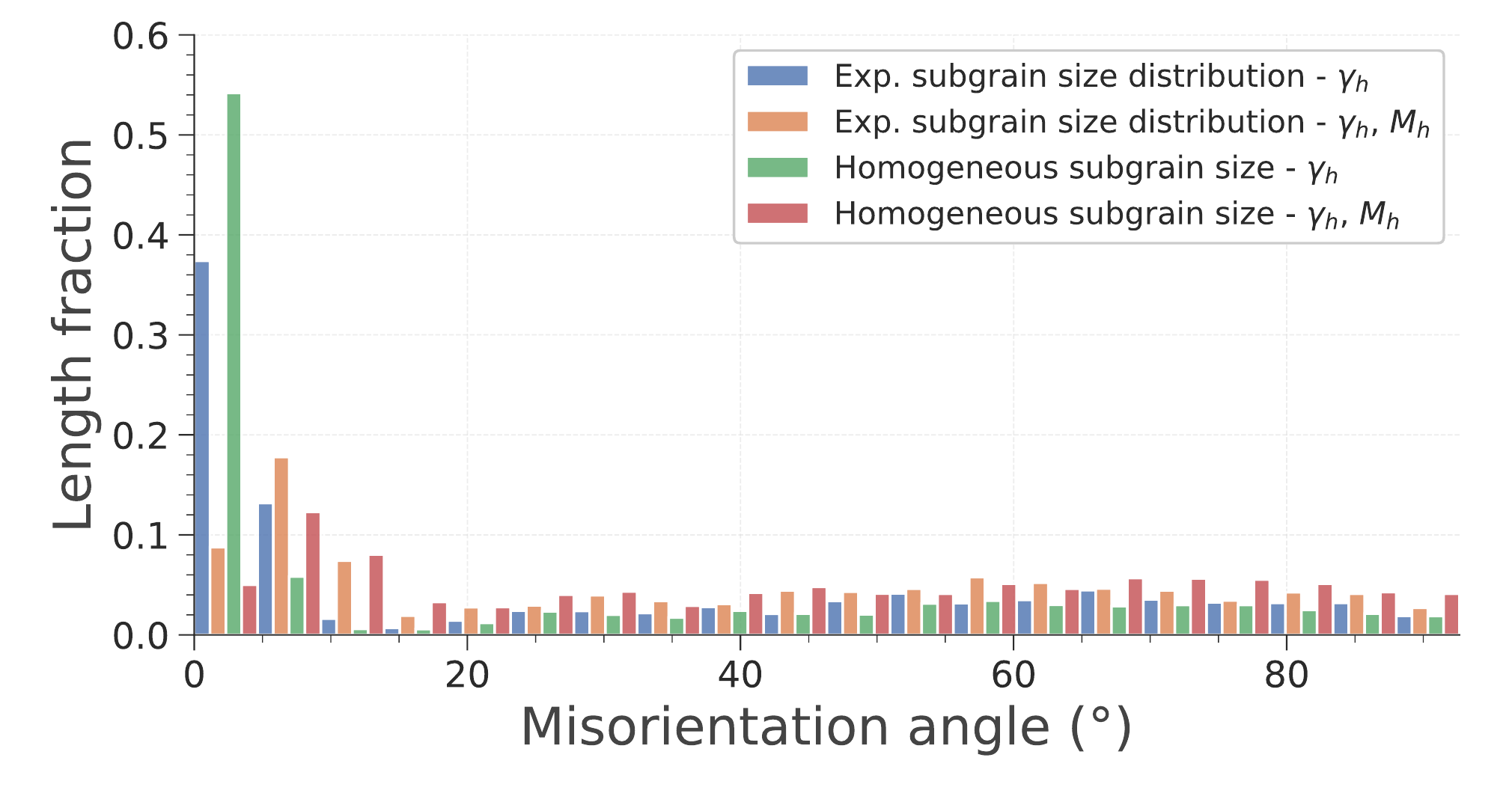}
        \caption{\label{fig:FinalDisorientationDistributions} Final state.}
    \end{subfigure}
\caption{Disorientation histograms at three instants of the simulation.}
\label{fig:DisorientationDistributions}
\end{figure}

Figure \ref{fig:GG_ECD} describes how the mean equivalent circle diameter (ECD) evolves with time during simulation. First, it is interesting to note that considering heterogeneous mobility (according to eq. \ref{eq:Mobility}) affects significantly subgrain growth and that its impact depends on initial subgrain size distribution. To assess this in more details, the evolution of total HAGB and LAGB length and of HAGB length fraction are plotted in figure \ref{fig:GBlengths}. HAGB length ratio (fig. \ref{fig:HAGBlengthratio}) exhibits how heterogeneous mobility favors the migration of HAGB and their formation by putting in contact subgrains originally located in different grains. At the end of the simulation, this leads to a higher HAGB length fraction. Moreover, as heterogeneous mobility favors HAGB migration, subgrains partially bounded by HAGB have greater chance to grow. This is particularly true in the case $d)$ of figure 5, in which all subgrains initially have the same size. Consequently, the advantage given by the high mobility of those boundaries is increased and their growth dominates the whole microstructure evolution. Figures \ref{fig:LAGBlength} and \ref{fig:HAGBlength} confirm those remarks by showing a temporary increase of HAGB total length if mobility is heterogeneous. Finally, it is interesting to note that if subgrains initially have the same diameter, an incubation time is needed for the subgrains partially closed by HAGB to take a significant advantage and start to consume the whole microstructure. Finally, figure \ref{fig:DisorientationDistributions} confirms these observations. It is interesting to note that both study cases assuming homogeneous mobility tends to keep a significant higher fraction of LAGB. This illustrates that a higher heterogeneity of GB induces a change in the behavior adopted to lower the total energy of the system. It changes from global reduction of both LAGB and HAGB length to temporary increase of HAGB length in order to speed up the decrease of total GB length.

These detailed observations confirm that the faster growth of a small fraction of subgrains is only observable if mobility is considered heterogeneous, as described by Holm \textit{et al.} \cite{Holm2003}. They also broaden the remarks made by Suwa \textit{et al.} \cite{Suwa2008} about the conditions allowing preferential subgrain growth. Indeed, this phenomenon is much less significant if initial subgrain size is not homogeneous and representative of experimental data.

\subsubsection{Influence of stored energy}\label{subsubsec:StoredEnergy}

In previous papers describing similar full-field simulations, driving pressure due to dislocation density heterogeneity is generally neglected \cite{Despres2020, Holm2003, Suwa2008}. This rests upon the simplification that deformation energy is fully consumed thanks to recovery, by annihilation and/or organization into GB network. The available framework allows us to assess the influence of the assumptions by considering that each subgrain has its own dislocation density. Two simulation cases are defined:
\begin{itemize}
    \item stored energy is initialized per subgrain by considering a dislocation density distribution taken from estimation of geometrically necessary dislocations (GND) by EBSD measurements.
    \item Stored energy is initialized per grain using the same distribution. Then, subgrain energy is initialized weighing the parent grain using coefficients (named $w_i$) respecting normal distribution. Parameters of this normal distribution are defined as follow: $w \in \left[ 0.1 ; 2.0 \right]$ ; $\overline{w} = 1.0$ ; $ \sigma^2 = 0.2$. Distribution parameters have been set as thereby to ensure subgrains inside grains do not have the exact same energy and that it is still a driving pressure.
\end{itemize}

\begin{figure}
    \begin{subfigure}{0.5\textwidth}
        \centering
        \includegraphics[width=0.95\linewidth]{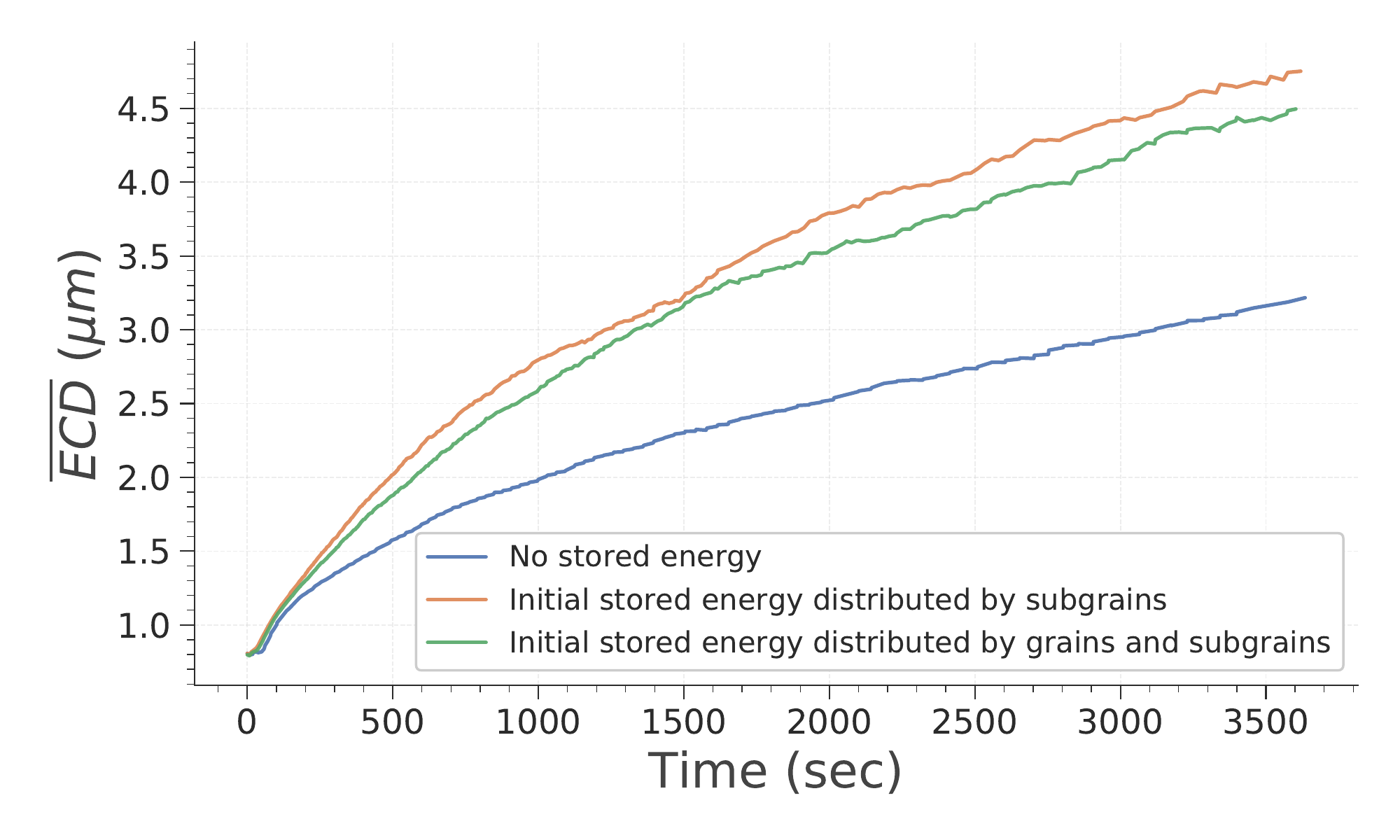}
        \caption{\label{fig:ECD_storedEnergy} Evolution of $\overline{ECD}$ as a function of time.}
    \end{subfigure}
    \begin{subfigure}{0.5\textwidth}
        \centering
        \includegraphics[width=0.95\linewidth]{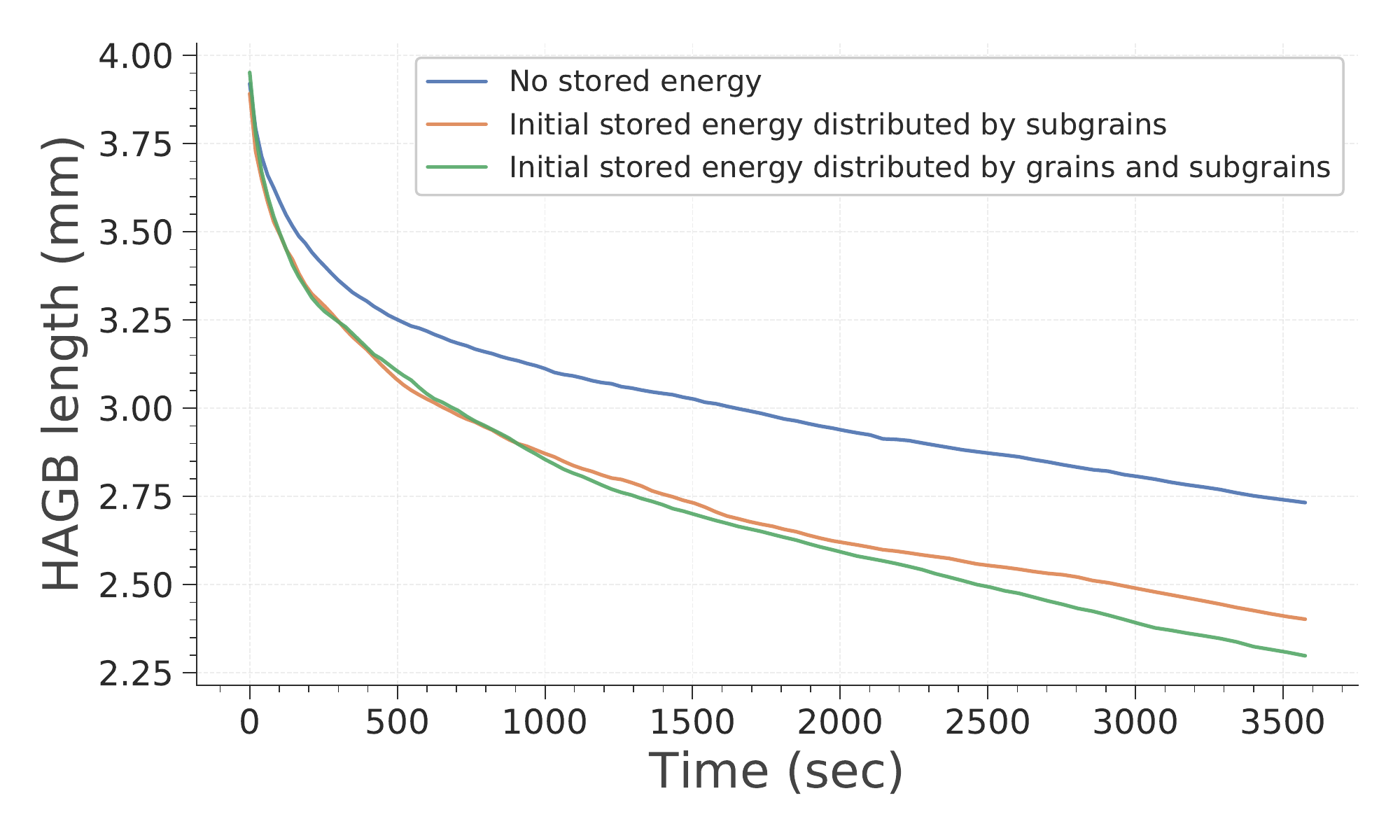}
        \caption{\label{fig:HAGBlength_storedEnergy} HAGB total length as a function of time.}
    \end{subfigure}
\caption{Main microstructural feature evolutions impacted by initial stored energy.}
\label{fig:ResultsStoredEnergy}
\end{figure}

Figure \ref{fig:ResultsStoredEnergy} presents the main microstructure features that are influenced by the presence of stored energy. As expected, considering stored energy variations between grains lead to an increase of the driving pressures and consequently to a faster growth of subgrains presenting an initial stored energy advantage. Since the total stored energy in the two last simulations is the same, differences in terms of kinetic are rather small. However, it is interesting to note that in the second case considering stored energy, subgrains and the complete microstructure evolve slightly slower whereas total HAGB length decreases faster. The faster disappearance of HAGB should be related to the stored energy distribution. Indeed, stored energy gaps between subgrains located at grain interfaces is higher than between subgrains belonging to the same parent grain. This contributes to speed up the movement of HAGB and to reduce the growth of subgrains located in inner grain regions.

\subsubsection{Discussion of numerical criterion for identification of recrystallized grains}\label{subsubsec:NumericalCriterionRXgrains}

In all of the papers presenting full-field simulations of microstructure evolution including substructures, recrystallized grains are identified based on their size. Subgrains are generally considered as recrystallized grains as soon as they are 6 to 8 times bigger than the initial mean subgrain size \cite{Despres2020, Holm2003, Suwa2008}. This relies on the hypothesis that all subgrains are free of energy and could become recrystallized grains. However, in recent experimental studies, recrystallized grains are generally identified based on some parameter quantifying internal disorientation or directly upon GND density \cite{ Rollett2017, Carneiro2020}. To correctly evaluate recrystallized fraction and being able to compare experimental to numerical data, several criteria are defined and compared. Subgrains are considered as recrystallized if:
\begin{itemize}
    \item they are 8 times bigger than the initial mean subgrain size,
    \item their internal dislocation density is lower than a given threshold,
    \item their internal dislocation density is lower than a given threshold and at least half of the boundaries surrounding it are HAGB.
\end{itemize}
Obviously, the last two criteria are only tested against the simulations that consider stored energy (see section \ref{subsubsec:StoredEnergy}).

\begin{figure}
	\centering
 	\includegraphics[width=0.65\textwidth]{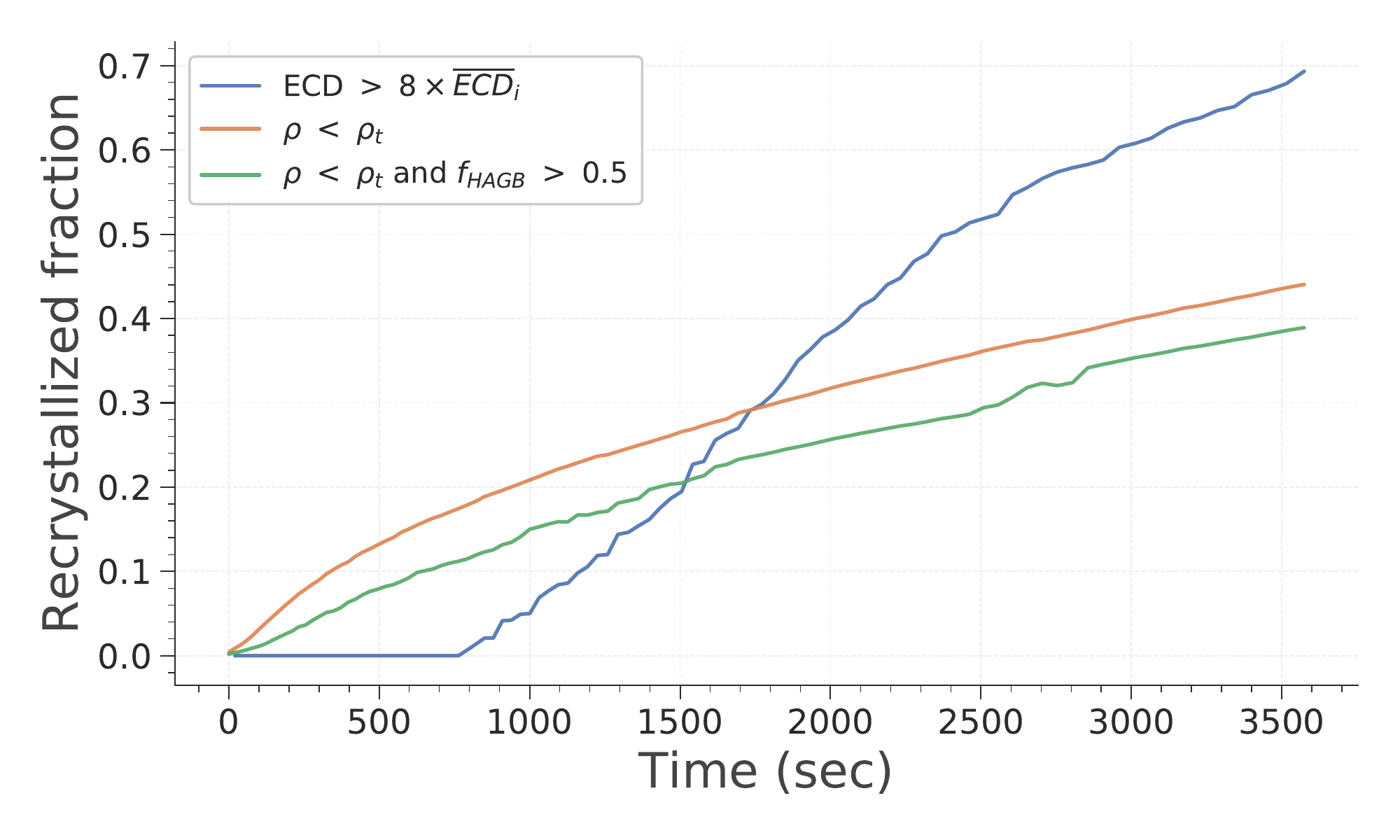}
  \caption{Evolution of recrystallized fractions with time.}\label{fig:ComparisonRecrystallizedCriteria}
\end{figure}

Figure \ref{fig:ComparisonRecrystallizedCriteria} illustrates how recrystallized surface fraction evolves during simulation, for the first simulation case described in section \ref{subsubsec:StoredEnergy}. As one could have expected, the recrystallized fraction behaves differently depending on its definition. Consequently, ensuring that the criteria used to discriminate recrystallized grains from experimental and simulation data are consistent is critical.

\subsection{Modeling of CDRX and PDRX}\label{subsec:CDRX}

Let's have a look now at simulations of CDRX and PDRX. To present the abilities of the model and evaluate the influence of subgrain formation rules, the results obtained with four different test cases are described below. All of them consider initial microstructures respecting the same grain size and hardening coefficient distributions. They include approximately 300 grains. The initial number of grains is taken low since it will increase substantially during deformation. Material parameters have been estimated based on the experimental results obtained conducting a thermomechanical testing campaign associated with extensive EBSD characterization. A detailed article will be dedicated to the presentation of these results and to the comparison with some simulation results acquired using this numerical framework. Thermomechanical conditions corresponding to these simulations are the following: $T = 650^{\circ}C ; ~ \dot{\varepsilon} = 1.0 ~ s^{-1} ; ~ \varepsilon_f = 1.35$.
The four test cases differ by the GB properties and the restrictions set for formation of subgrains. In all of the cases, except the last ones, subgrains cannot be placed too close to grain boundaries to avoid topological events that are not seen experimentally such as bulging boundaries. They are defined as followed:
\begin{description}
    \item[(a)] The number of subgrains that are formed at each deformation increment is computed individually per grain/subgrain. GB energy is described by RS equation and GB mobility is isotropic.
     \item[(b)]  The number of subgrains that are formed at each deformation increment is computed for the whole domain. Then, new subgrains are positioned randomly within the RVE. GB energy is described by RS equation (Eq. \ref{eq:GammaRS}) and GB mobility is isotropic. 
     \item[(c)]  The number of subgrains that are formed at each deformation increment is computed individually per grain/subgrain. GB energy is described by RS equation (Eq. \ref{eq:GammaRS}) and GB mobility is heterogeneous and computed using equation \ref{eq:Mobility}.
    \item[(d)]  This last case respects the same rules than the first one, except that new subgrains can be placed on pre-existing grain boundaries.
\end{description}

Based on the discussion presented previously (see section \ref{subsubsec:NumericalCriterionRXgrains}) and to be as close as possible to the criterion used commonly when working with experimental data, it has been decided to define as recrystallized the grains that fulfill the two following criteria:
\begin{itemize}
    \item $\rho \le \rho_{th} = 1.0 \times 10^{14} ~ m^{-2} $,
    \item only grains, i.e. entities bounded by at least $50 \%$ of HAGB, are included into the measure of recrystallized grains.
\end{itemize}

\subsubsection{Evolution of main microstructure descriptors during CDRX and PDRX}\label{subsubsec:EvolutionMainMicrostructureDescriptors}

\begin{figure}
    \begin{subfigure}{0.5\textwidth}
        \centering
        \includegraphics[width=0.95\linewidth]{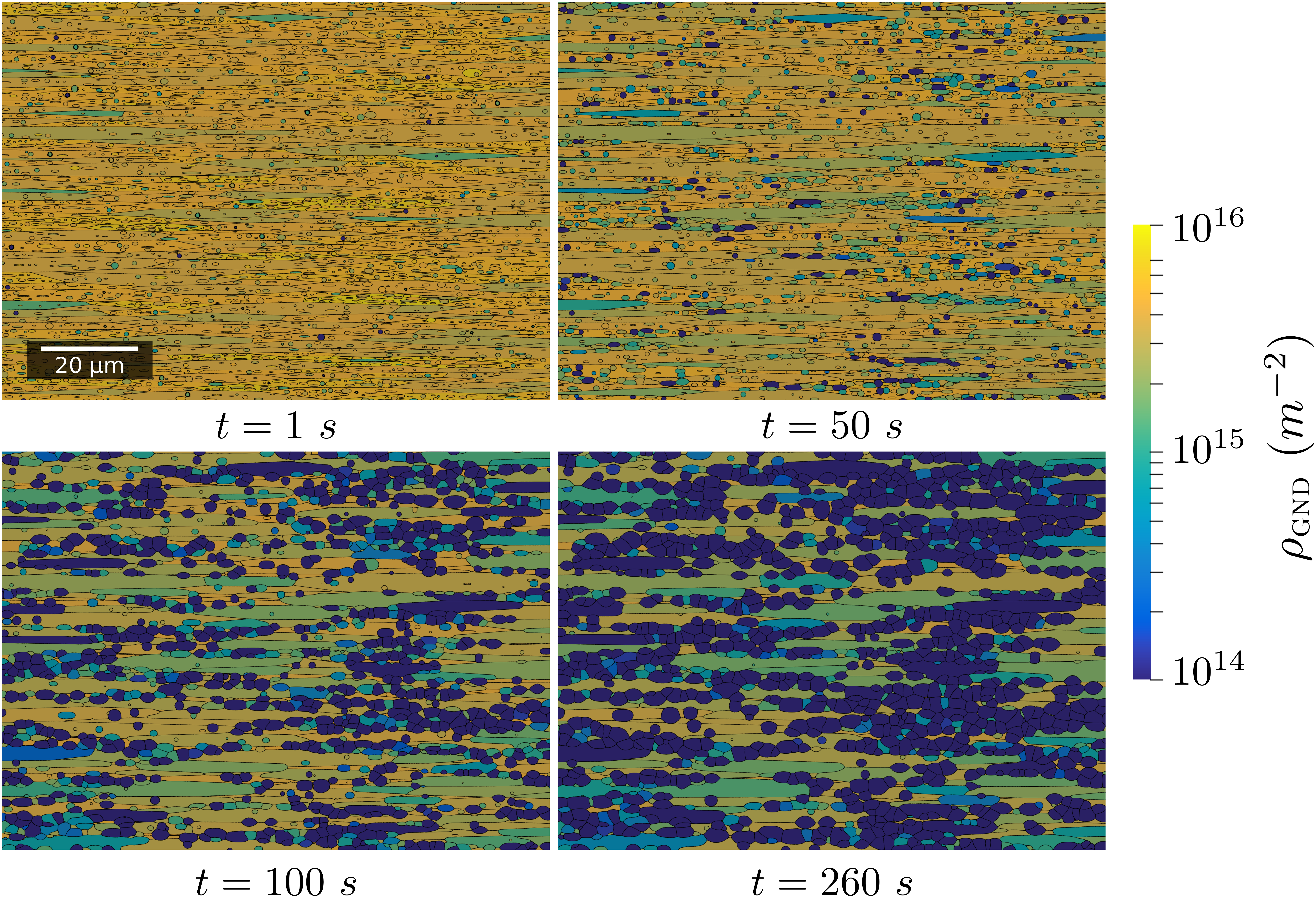}
        \caption{\label{fig:CDRX_LocalNucleation} Local nucleation - $\gamma_h$.}
    \end{subfigure}
    \begin{subfigure}{0.5\textwidth}
        \centering
        \includegraphics[width=0.95\linewidth]{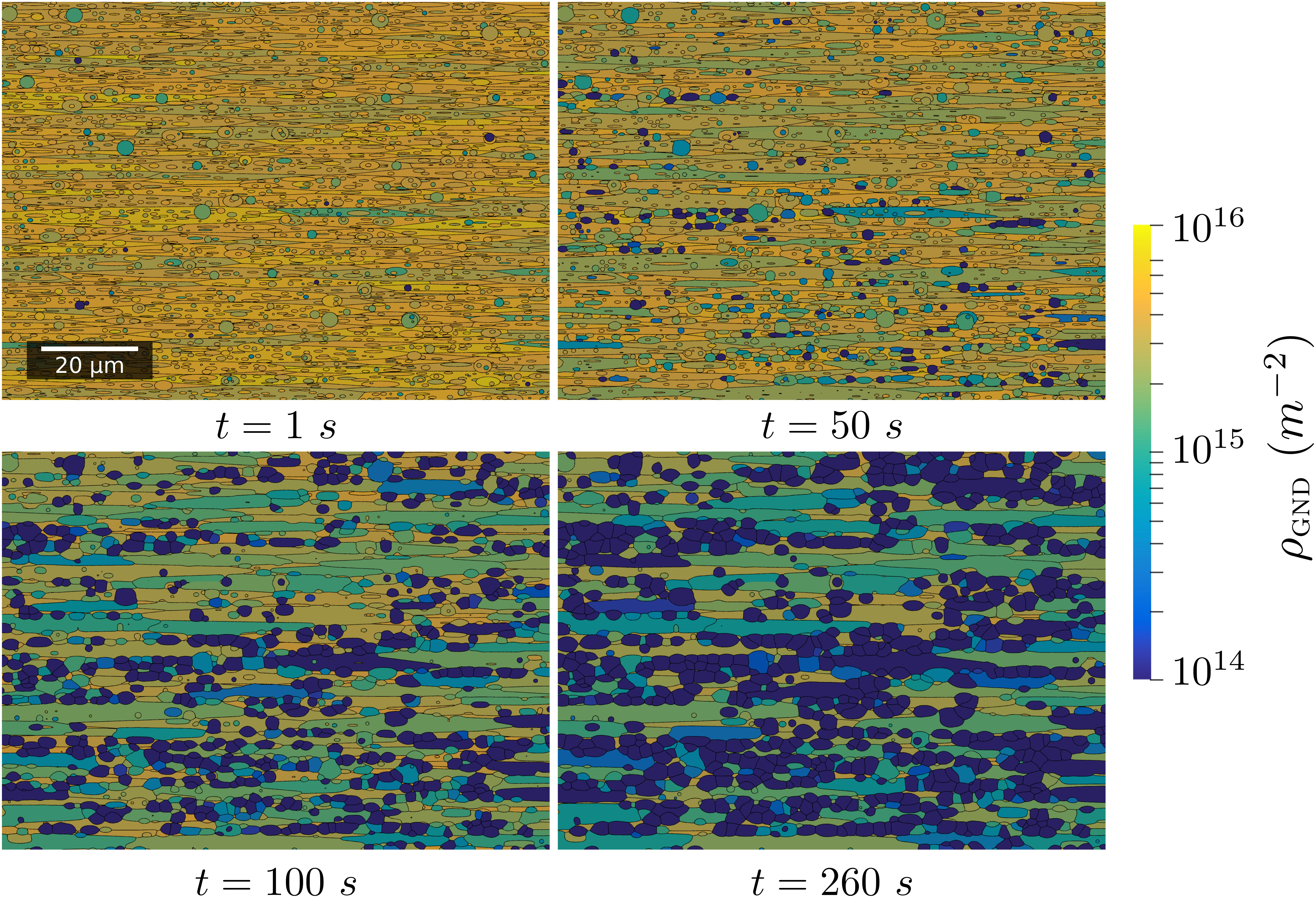}
        \caption{\label{fig:CDRX_GlobalNucleation} Global nucleation - $\gamma_h$.}
    \end{subfigure}\vspace{0.3cm}
    \begin{subfigure}{0.5\textwidth}
        \centering
        \includegraphics[width=0.95\linewidth]{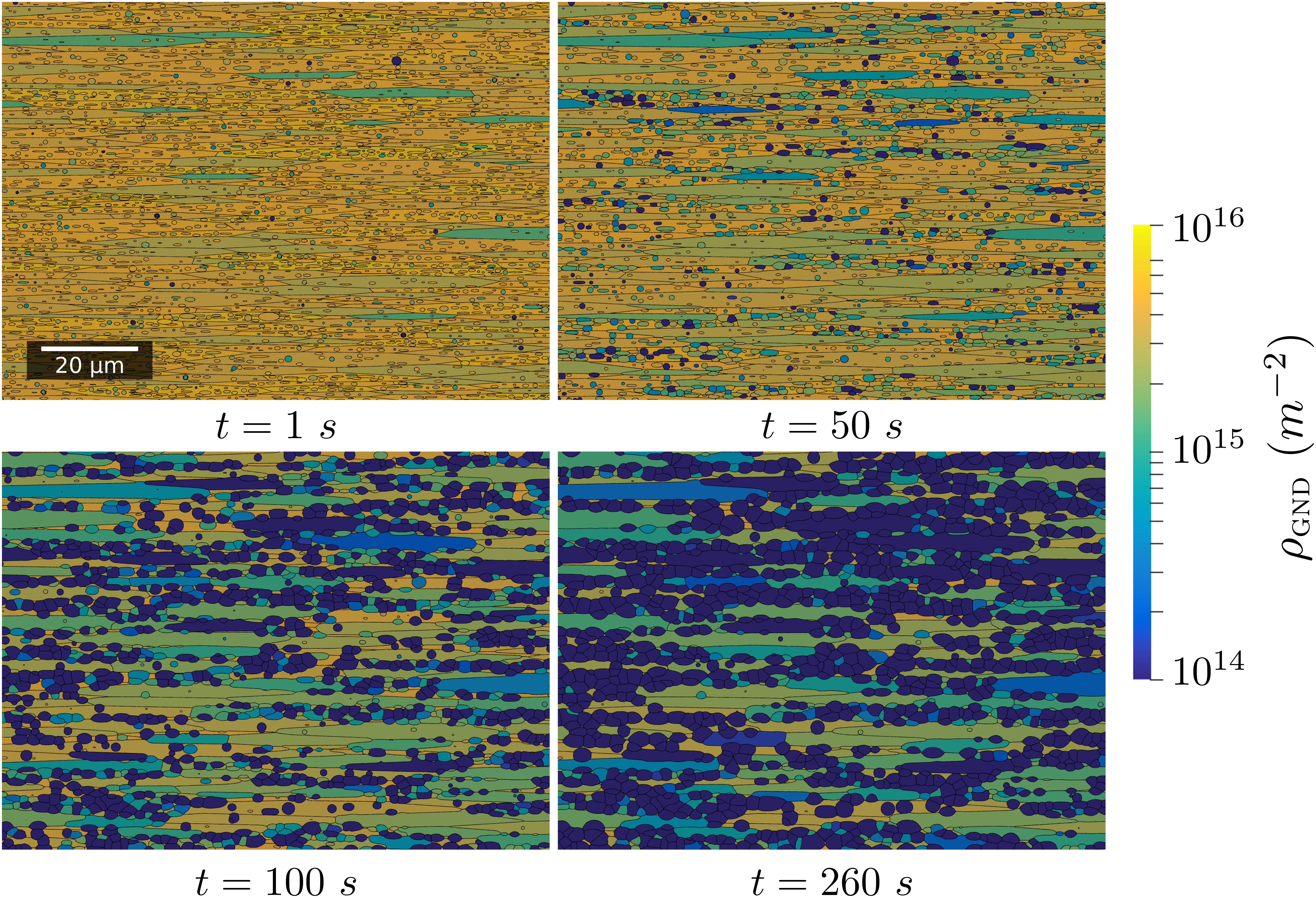}
        \caption{\label{fig:CDRX_HeterogeneousMobility} Local nucleation - $\gamma_h, ~ M_h$.}
    \end{subfigure}
    \begin{subfigure}{0.5\textwidth}
        \centering
        \includegraphics[width=0.95\linewidth]{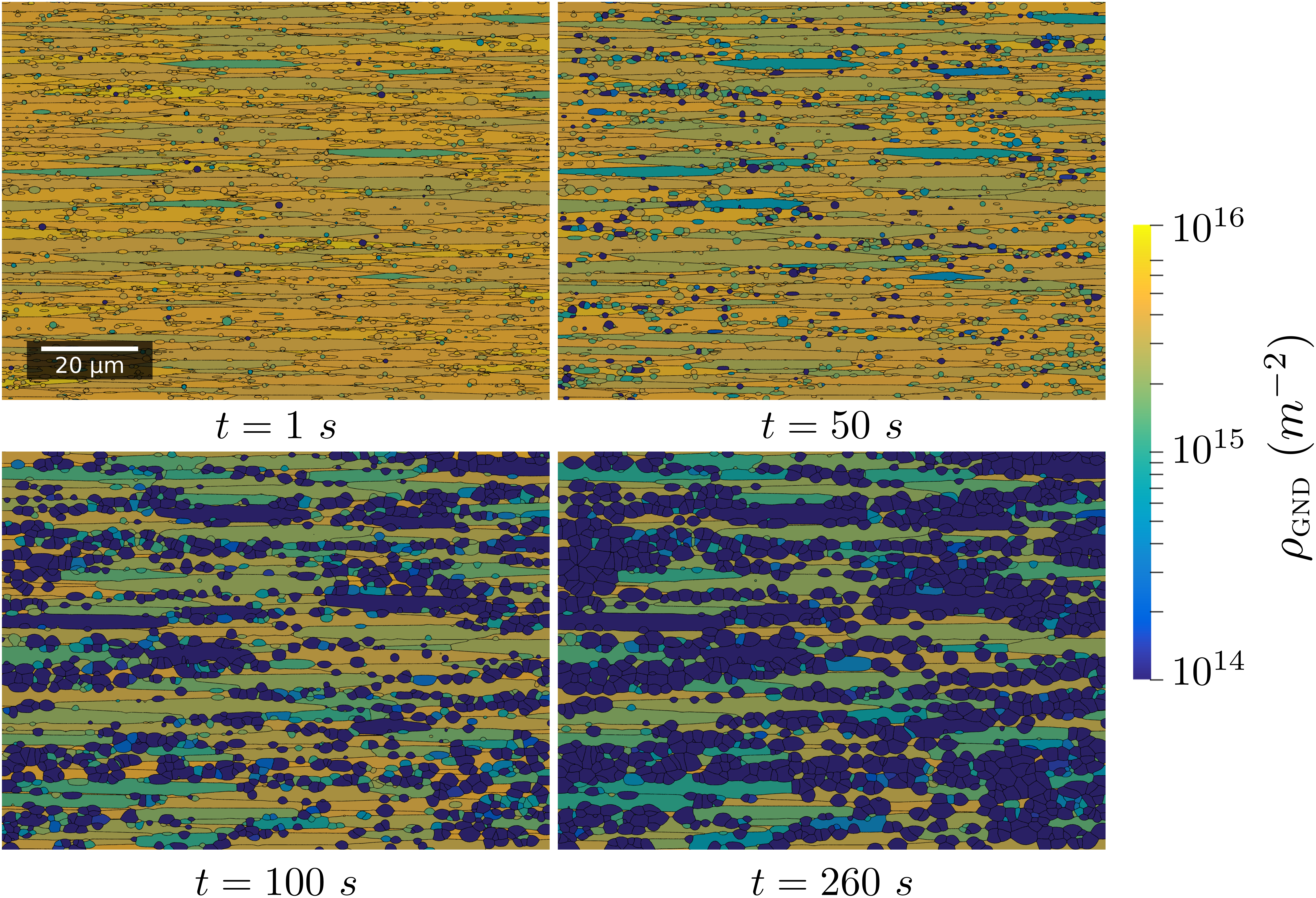}
        \caption{\label{fig:CDRX_NoIndicator} Local nucleation - $\gamma_h, ~ d_{NewLAGB}^{GB}=0$.}
    \end{subfigure}
\caption{Evolution of digital microstructures with time for the four different test cases.}
\label{fig:CDRX_microstructureEvolution}
\end{figure}

\begin{figure}[h!]
    \begin{subfigure}{0.5\textwidth}
        \centering
        \includegraphics[width=0.95\linewidth]{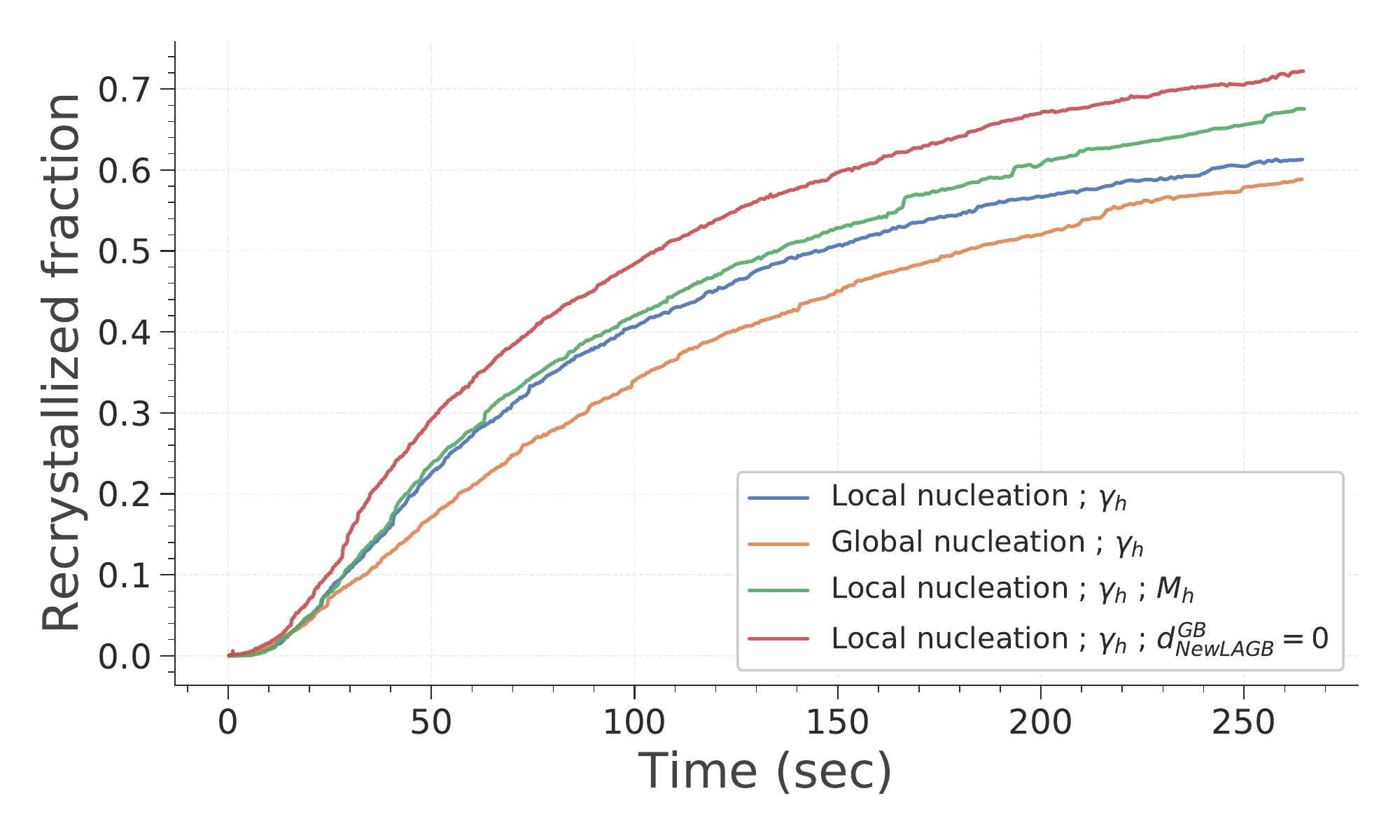}
        \caption{\label{fig:CDRX_RxFraction} Recrystallized surface fraction.}
    \end{subfigure}
        \begin{subfigure}{0.5\textwidth}
        \centering
        \includegraphics[width=0.95\linewidth]{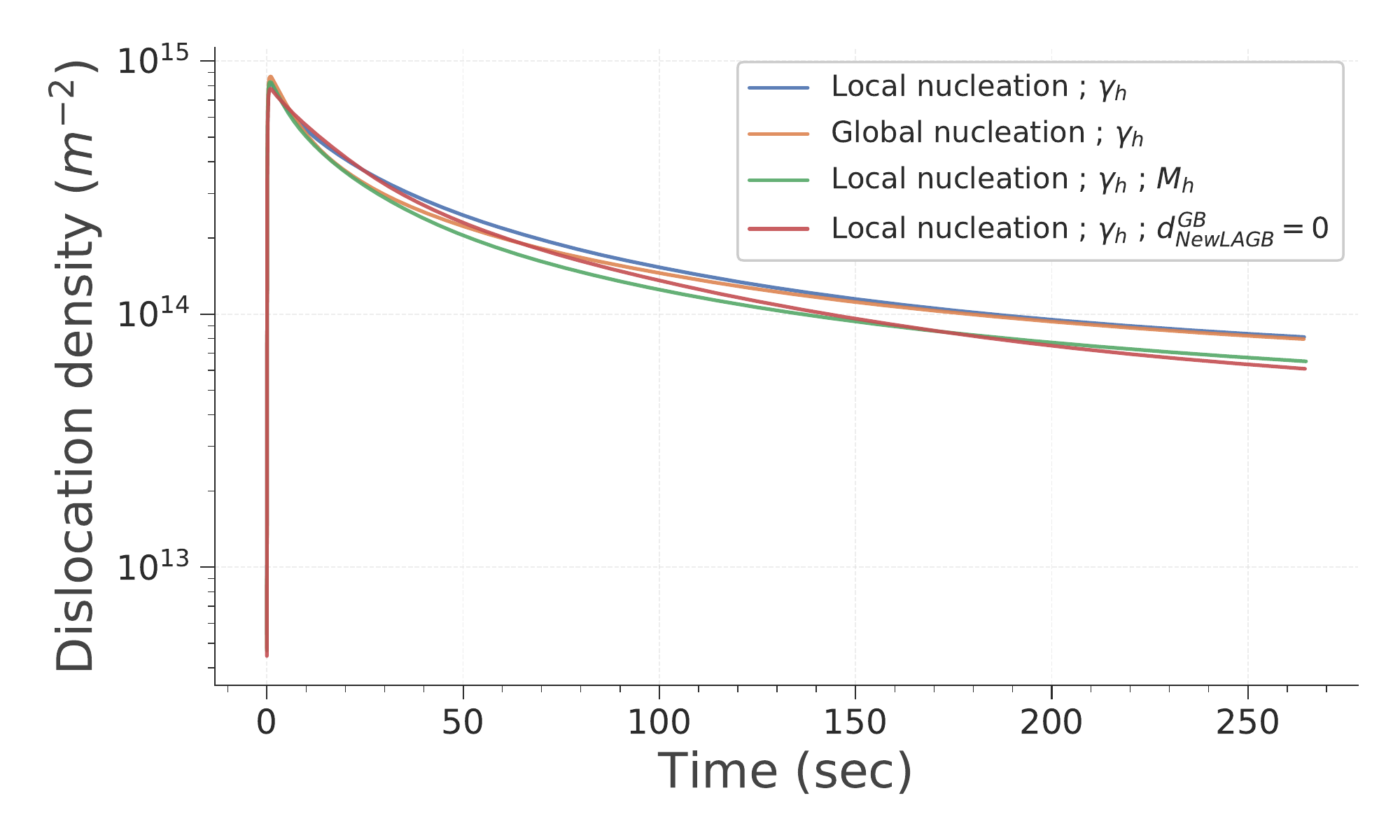}
        \caption{\label{fig:CDRX_GNDdensity} GND density.}
    \end{subfigure}\vspace{0.3cm}
    \begin{subfigure}{0.5\textwidth}
        \centering
        \includegraphics[width=0.95\linewidth]{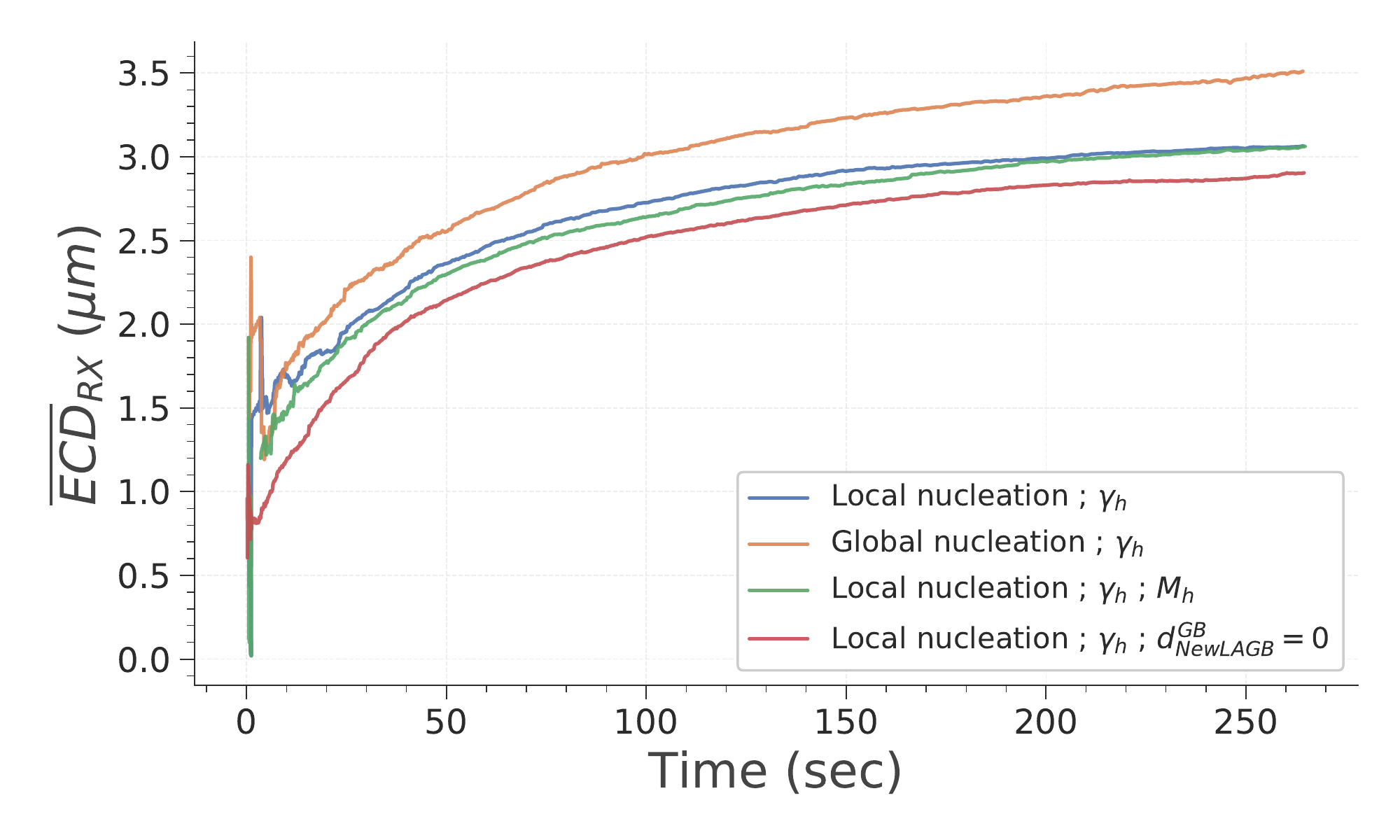}
        \caption{\label{fig:CDRX_ECD} Mean recrystallized grain size.}
    \end{subfigure}
    \begin{subfigure}{0.5\textwidth}
        \centering
        \includegraphics[width=0.95\linewidth]{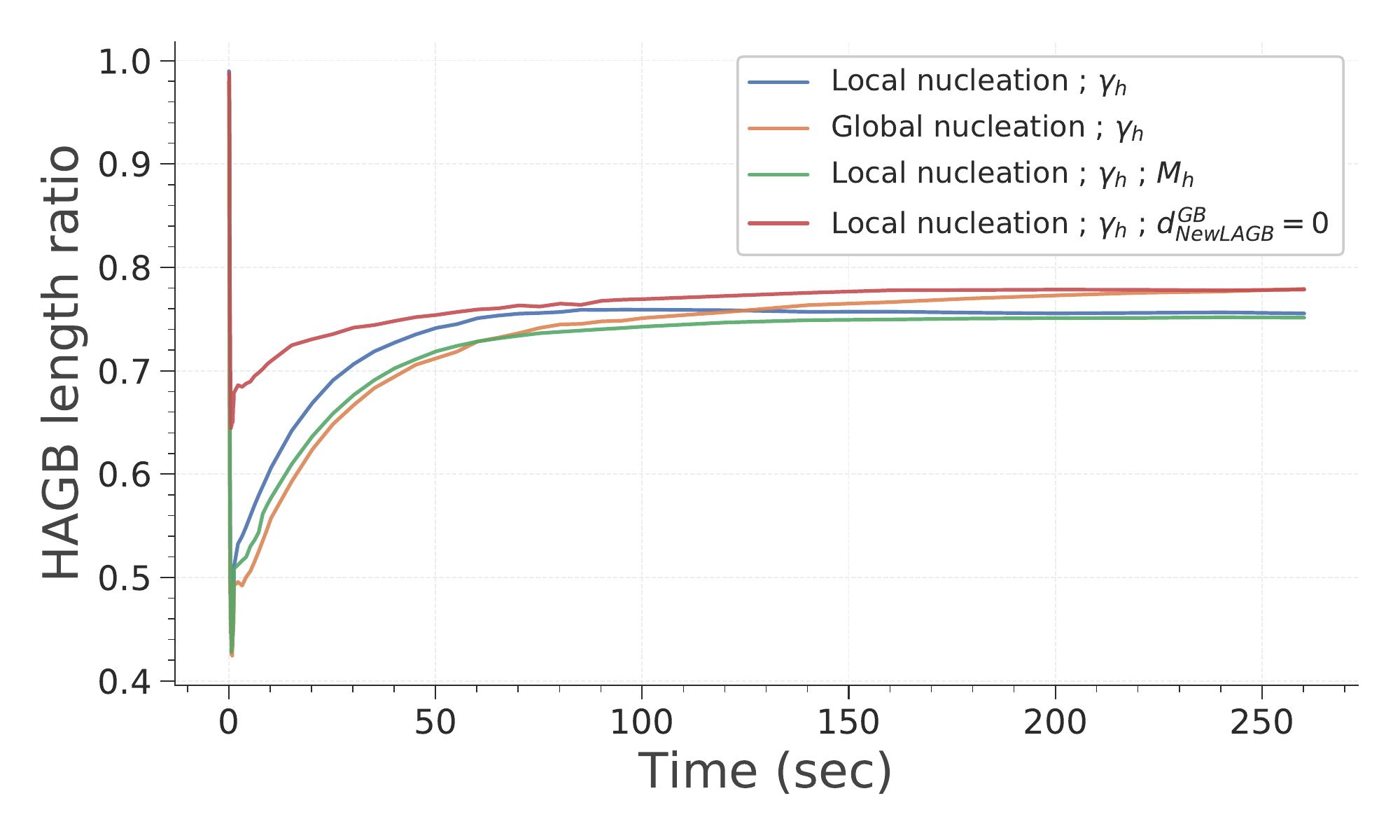}
        \caption{\label{fig:CDRX_HAGBfraction} HAGB length ratio.}
    \end{subfigure}
\caption{Main microstructural feature evolution with time.}
\label{fig:CDRX_mainFeatures}
\end{figure}

Figure \ref{fig:CDRX_microstructureEvolution} presents evolution of digital microstructures for the four test cases, after a deformation of $1.35$ and a subsequent holding at temperature (hundreds of seconds at $650 ^{\circ}C$). Figure \ref{fig:CDRX_mainFeatures} presents the evolution of the main microstructural descriptors with time. First, one could see that the recrystallized fraction is null during the deformation process and starts to increase significantly only after tens of seconds of holding at temperature. This can be explained easily by analyzing the criteria defined previously for discrimination of recrystallized grains in regards of the mechanisms introduced into the numerical framework. First, only subgrains are formed during deformation due to recovery of a certain fraction of dislocations. These objects do not fulfill the second criterion. Then, under subsequent deformation, their misorientation increase as prescribed by equation \ref{eq:ProgressiveMisorientation}. However, at the same time, their internal dislocation density increases. Therefore, with the material parameters considered here, new subgrains see their dislocation density exceed the threshold value before they have time to transform into grains. Then, under holding at temperature, grains and subgrains that have an energetic advantage over the surrounding grains start to grow. Their dislocation density is progressively lowered. They are also likely to encounter new grains which generally lead to an increase of the bounding HAGB fraction. Consequently, more and more candidates are meeting both of the criteria and are flagged as recrystallized. Finally, evolution of average recrystallized grain size presents a sigmoidal shape, except for the first instants where there are abrupt and erratic variations. This is due to the small number of grains considered as recrystallized at those times. Finally, it is interesting to discuss how HAGB length ratio evolves during the whole process (figure \ref{fig:CDRX_HAGBfraction}). First, it decreases quickly with strain due to the fact that some new subgrains form. Then, once deformation is stopped, it increases by approximately $10 \%$ instantaneously. One reason for that observation is that the hardening parameter, $K_1$, attributed to each subgrain is taken from a distribution. Therefore, some subgrains have a higher hardening parameters than the parent grains and disappear as soon as their dislocation density is higher than the one of the parent grain. This is not visible during deformation since the formation of new subgrains is much more significant. Secondly, it is noteworthy to mention the rapid stagnation of HAGB length ratio. This would mean that grains and subgrains that have a lower dislocation density grow with similar rates. It makes sense since the driving pressure induced by interface energy is negligible in regards of the one linked to differences in deformation stored energy.

Considering now the differences between the results obtained with the four test cases, it appears that considering a heterogeneous GB mobility does not impact significantly the statistical descriptors described here. One could also notice that introducing subgrains over the whole domain globally affect recrystallized fraction at intermediate times and induces an increase of the average recrystallized grain size by approximately $15 \%$. This difference in recrystallized fraction can be explained by the fact that the main driving pressure for the growth of recrystallized grains is the difference in stored energy between those grains and the surrounding ones. Therefore, if the quantity of interfaces created during deformation is individualized per grain, one can deduce from equation \ref{eq:SurfaceNewSubgrains} that more subgrains will be placed in zones with high dislocation density. Thus, the kinetic of recrystallization will be higher for intermediate holding times. One factor in favor of the difference in  average recrystallized grain size is that, in the case with global nucleation, viable subgrains that effectively form recrystallized grains can grow much significantly before encountering other recrystallized grains. Finally, regarding the last test case, allowing new subgrains to form at GB leads to $10 \%$ higher recrystallized fraction. The reason behind that difference is that subgrains that form over pre-existing GB have already a fraction of their boundary as HAGB.

\subsubsection{Evolution of the subgrain network during deformation}\label{subsubsec:EvolutionSubgrainNetworkDeformation}

\begin{figure}[h!]
    \begin{subfigure}{0.5\textwidth}
        \centering
        \includegraphics[width=0.95\linewidth]{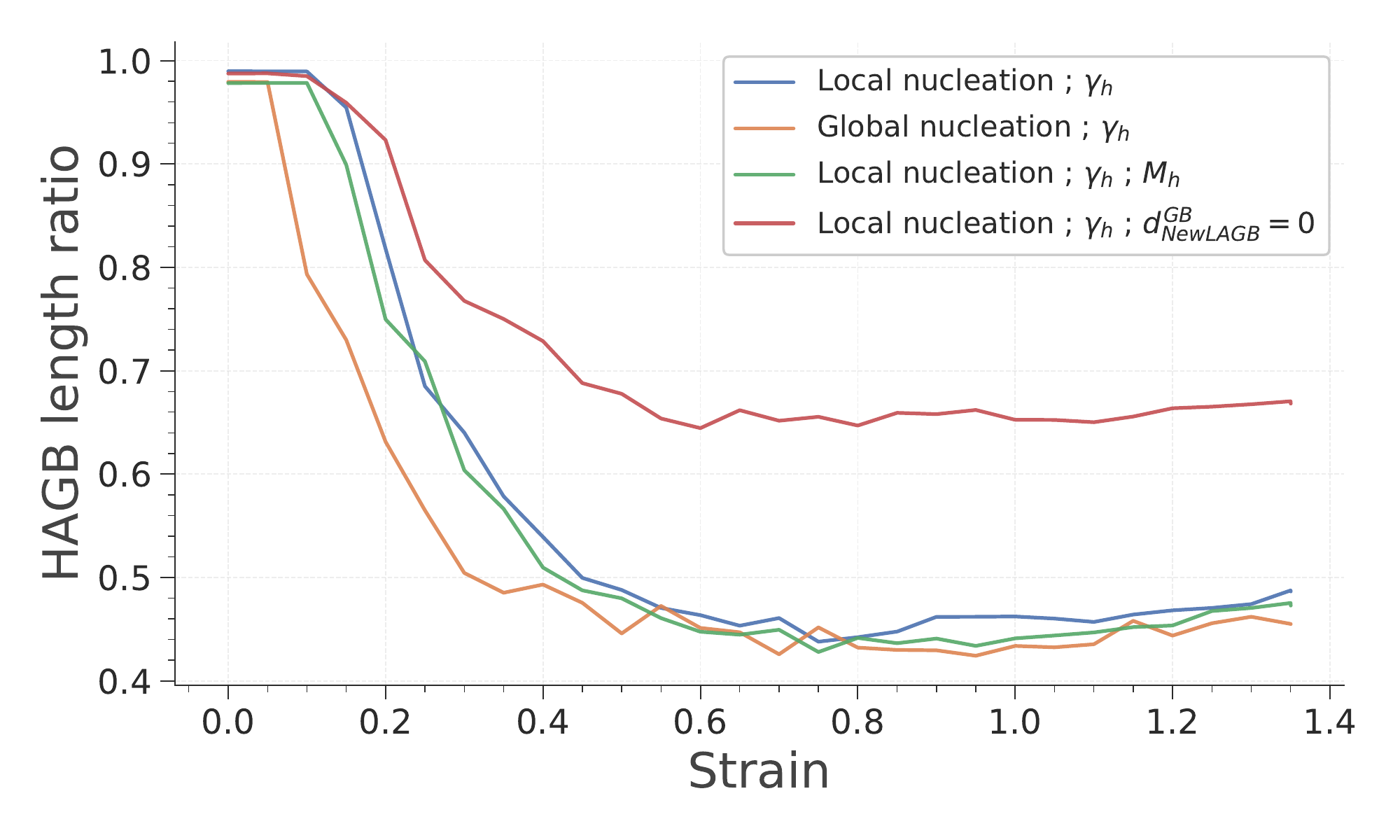}
        \caption{\label{fig:CDRX_HAGBlengthRatio_Strain} HAGB length ratio.}
    \end{subfigure}
        \begin{subfigure}{0.5\textwidth}
        \centering
        \includegraphics[width=0.95\linewidth]{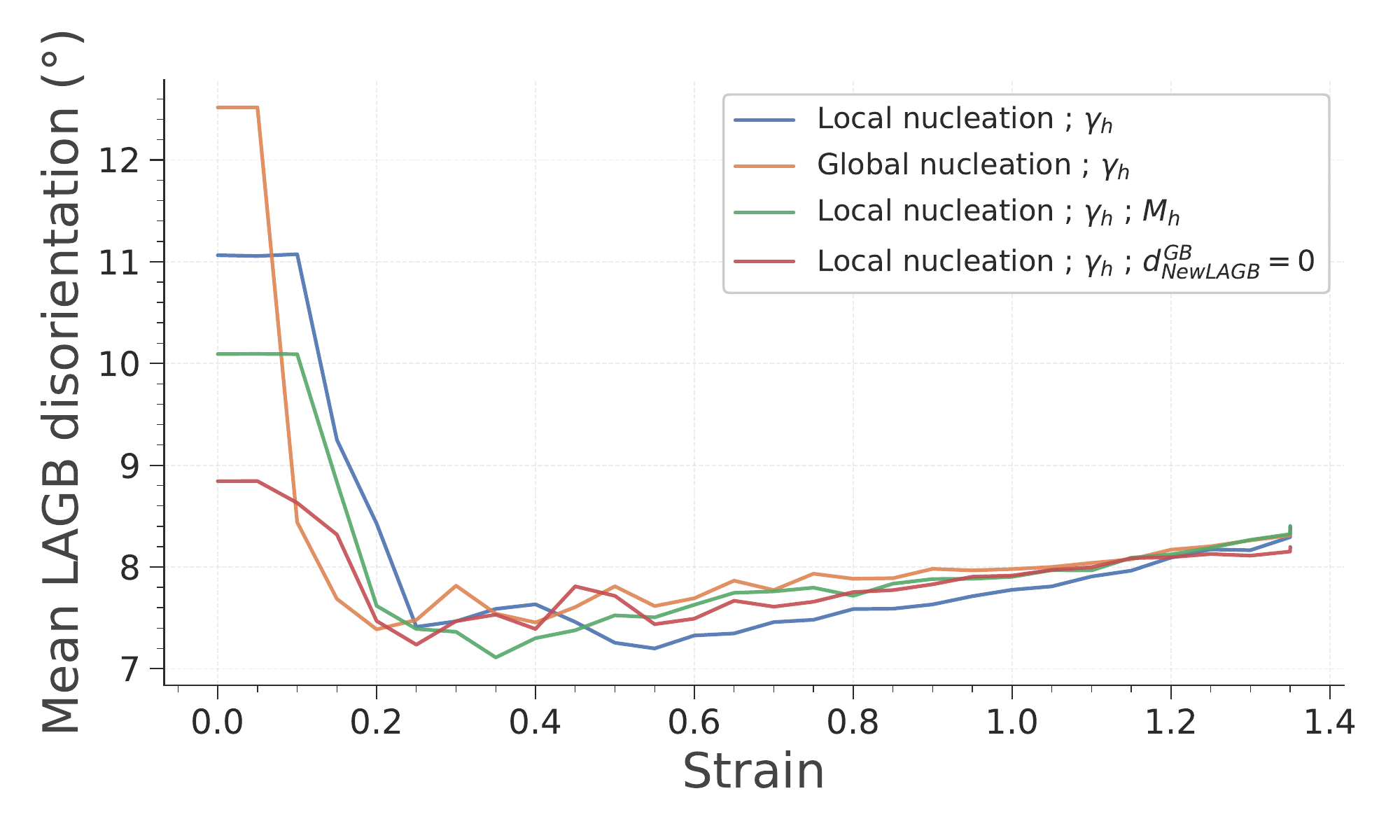}
        \caption{\label{fig:CDRX_MeanLAGBdisorientation_Strain} Mean LAGB disorientation.}
    \end{subfigure}
\caption{Main microstructural feature evolution with strain, during deformation.}
\label{fig:CDRX_mainFeaturesDeformation}
\end{figure}

The evolution of the HAGB length ratio and of the mean LAGB disorientation with strain are presented in figure \ref{fig:CDRX_mainFeaturesDeformation}. Focusing on HAGB length ratio first, it is interesting to note that it reaches a steady state after a strain of $0.6$. This is consistent with some experimental observations reported by Chauvy \textit{et al.} \cite{Chauvy2006}. As discussed in previous section, some subgrains disappear during deformation since they harden too much to stay viable. The average LAGB disorientation increases by approximately $1 ^{\circ}$ during deformation. It is a rather low increase, however it is important to note that new subgrains continue to form whereas some other shrink. Therefore, this evolution isn't fully representative of the disorientation increase due to equation \ref{eq:ProgressiveMisorientation}.

\begin{figure}[h!]
	\centering
 	\includegraphics[width=0.65\textwidth]{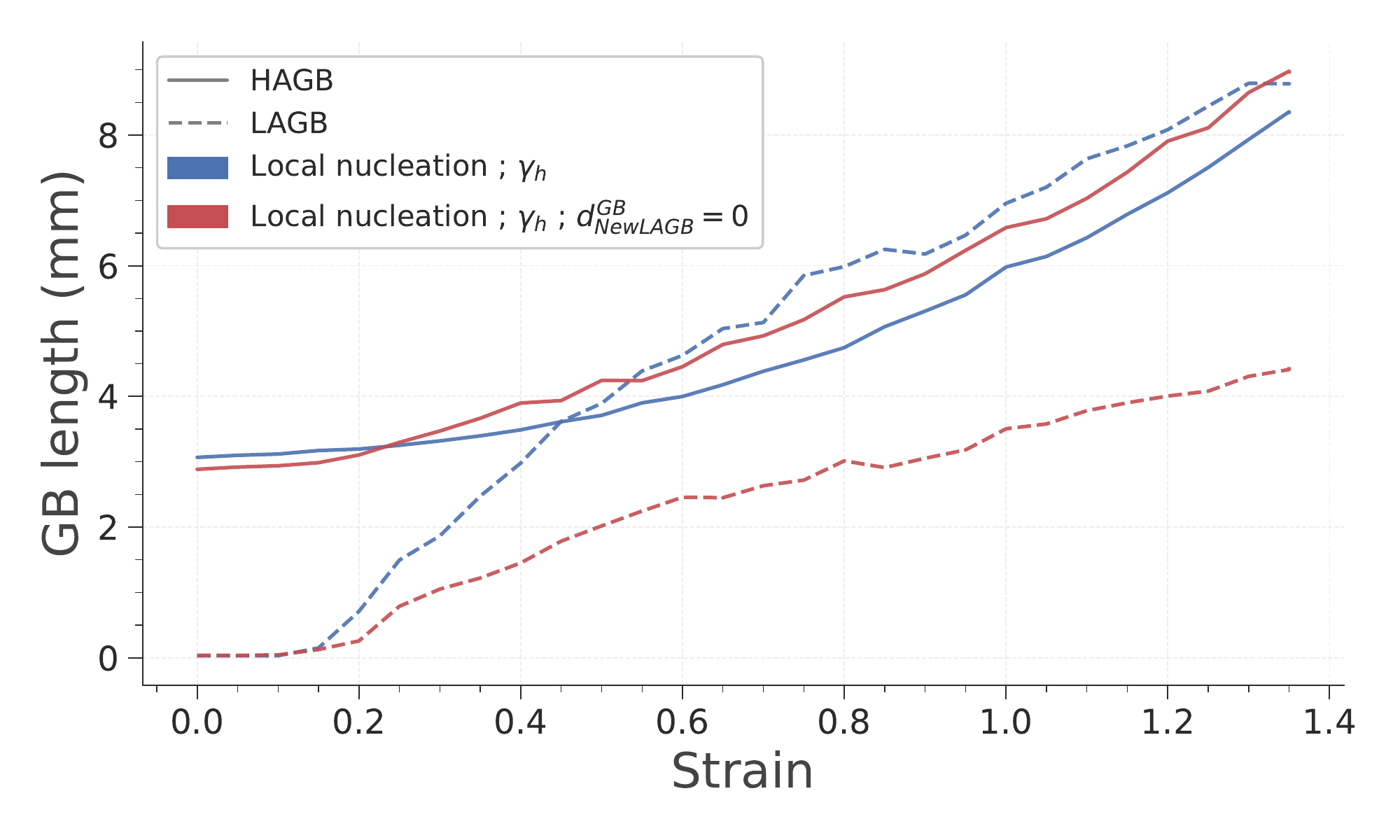}
  \caption{Evolution of LAGB and HAGB length with strain during deformation.}\label{fig:CDRX_inDetailAnalysisGBnetwork}
\end{figure}

 Finally, the last worth mentioning point is that the only case that exhibits different results regarding those two features is the last one, in which subgrains can form onto GB. To investigate the reason behind this difference in HAGB length ratio, the evolution of LAGB and HAGB length with strain is plotted in figure \ref{fig:CDRX_inDetailAnalysisGBnetwork}. It is interesting to note that both cases exhibit comparable HAGB length evolution. However, it appears that the reference case in which subgrains form at a given distance from GB, the increase of LAGB length is much higher. This is totally consistent since the formation of subgrains over GB gives rise to less low angle interfaces and at the same time the disappearing of HAGB.

\section{Conclusion}\label{sec:Conclusion}

The present study has detailed the current abilities of the proposed numerical framework to predict the evolution of subgrain structures by GG, CDRX and PDRX phenomena. The results concerning GG of a fully substructured microstructure has shone a light on:
\begin{itemize}
    \item the influence of the microstructure topology,
    \item how significant is the definition of LAGB and subgrain properties. Indeed, it appeared that the strategy adopted by the microstructure to reduce the total system energy is different depending on those parameters. The observations have led to deduce that heterogeneous mobility encourages a temporary increase of HAGB length to speed up the general energy decrease. They also illustrated that preferential subgrain growth is much more significant if subgrain all have initially the same size.
    \item The fact that internal dislocation density has a non negligible impact.
\end{itemize}
The results regarding simulation of CDRX and PDRX have illustrated that this simulation environment coupled to the Gourdet-Montheillet model provides realistic and consistent results. They also show that the model is flexible enough to assess the impact of several hypotheses and assumptions.

The results of the present CDRX model now need to be confronted to experimental results to assess to which extent they are able to predict them. This will be presented in an upcoming article regarding microstructure evolution of zircaloy-4 in hot forming conditions.
Moreover, the present CDRX and PDRX model could still be extended by being applied to other materials which exhibit different microstructure features due to the same CDRX mechanisms.

\vspace{6pt}



\section*{Author contributions}
Conceptualization, V.G., B.F., M.B.; software, V.G., M.B.; validation, V.G.; formal analysis, V.G.; investigation, V.G.; resources, M.B., A.G.; project administration, M.B., A.G.; funding acquisition, M.B., A.G.; writing—original draft preparation, V.G.; writing—review and editing, V.G., B.F., A.G., M.B.; supervision, A.G., M.B.. All authors have read and agreed to the published version of the manuscript.

\section*{Data availability}
The data necessary to reproduce these findings are available from the corresponding author on request.

\section*{Conflicts of interest}
The authors declare no conflict of interest. The funders had no role in the design of the study; in the collection, analyses, or interpretation of data; in the writing of the manuscript; or in the decision to publish the results.

\clearpage
\appendix

\section{Evolution of the GB network in the whole simulated RVE}\label{app:EvolutionGBnetworkFullRVE}

\begin{figure}[h!]
    \begin{subfigure}[t]{0.5\textwidth}
        \centering
        \includegraphics[width=0.95\linewidth]{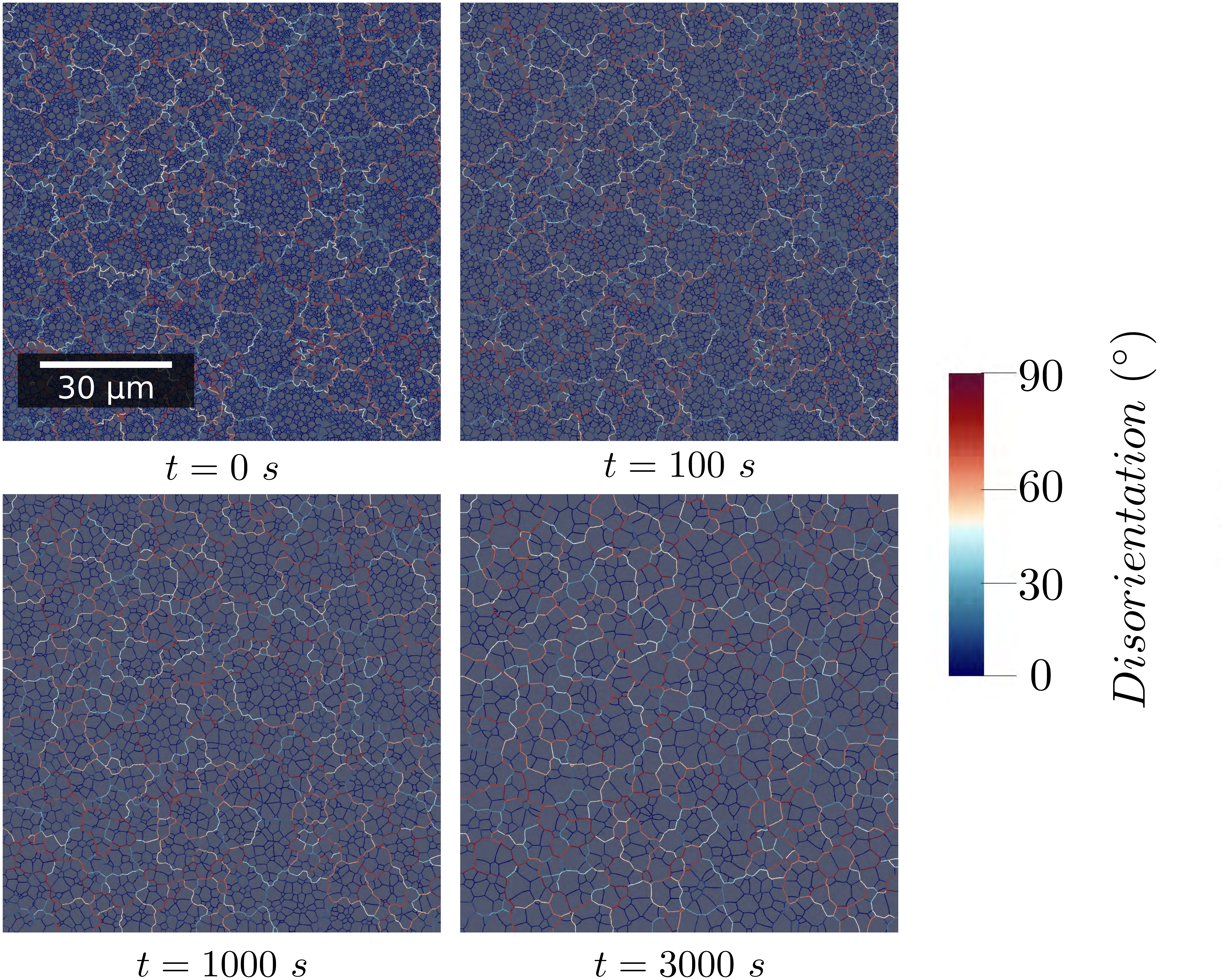}
        \caption{\label{fig:GG_HetGamma_FullRVE} Experimental subgrain size distribution - $\gamma_h$.}
    \end{subfigure}
    \begin{subfigure}[t]{0.5\textwidth}
        \centering
        \includegraphics[width=0.95\linewidth]{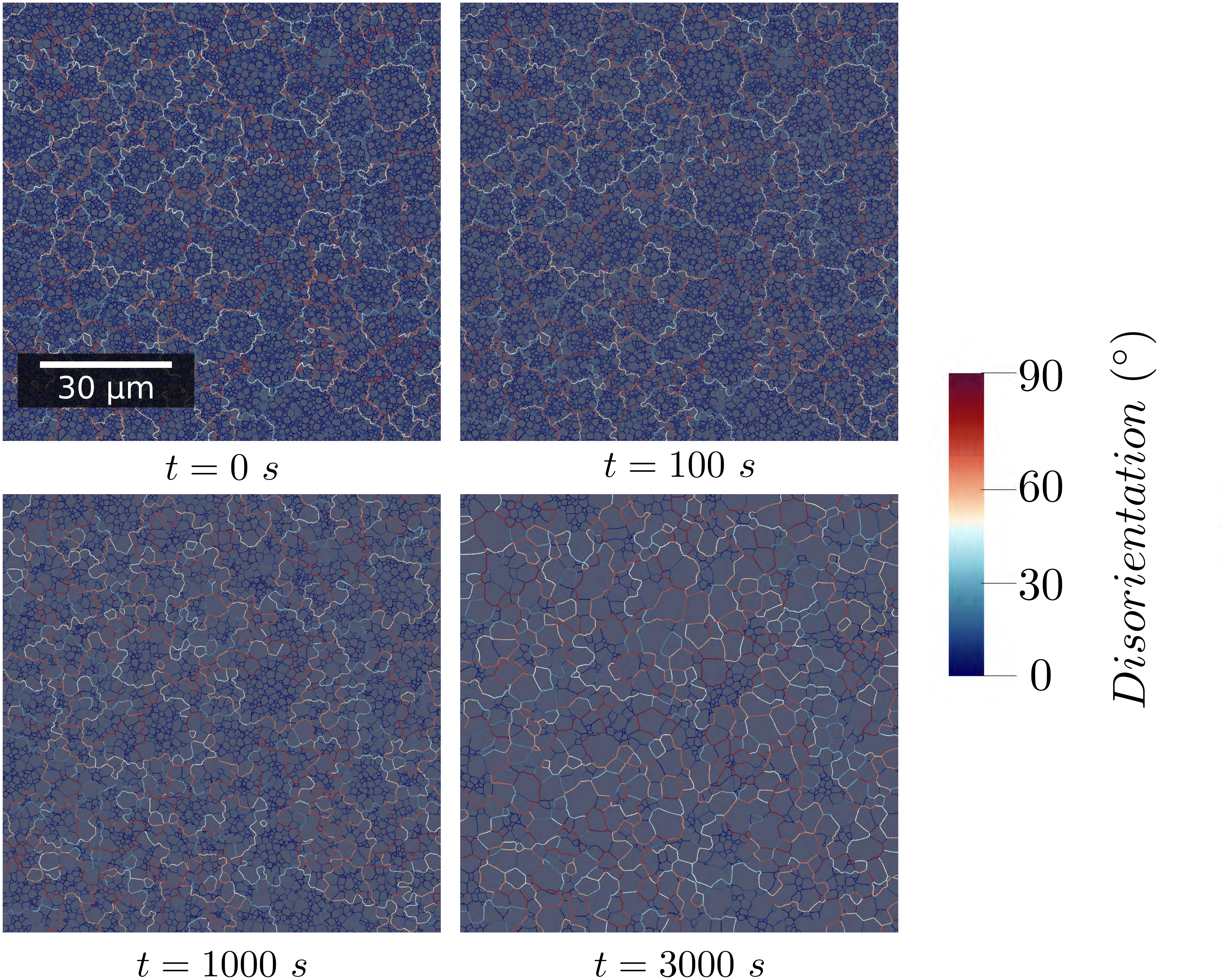}
        \caption{\label{fig:GG_HetGammaHetMob_FullRVE} Experimental subgrain size distribution - $\gamma_h, ~ M_h$.}
    \end{subfigure}\vspace{0.3cm}
    \begin{subfigure}[t]{0.5\textwidth}
        \centering
        \includegraphics[width=0.95\linewidth]{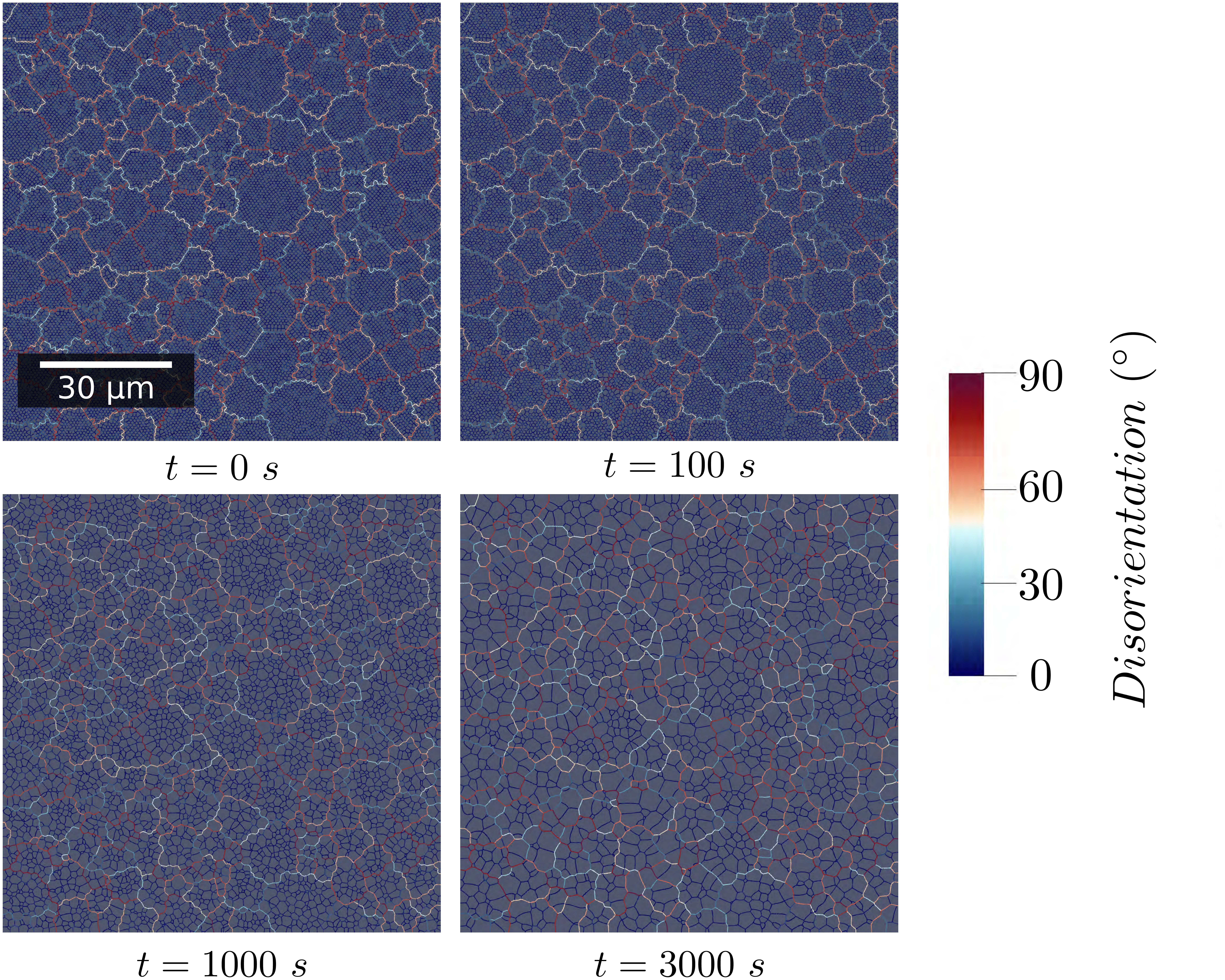}
        \caption{\label{fig:GG_HetGammaUniformSubgrainSize_FullRVE} Uniform subgrain size distribution - $\gamma_h$.}
    \end{subfigure}
    \begin{subfigure}[t]{0.5\textwidth}
        \centering
        \includegraphics[width=0.95\linewidth]{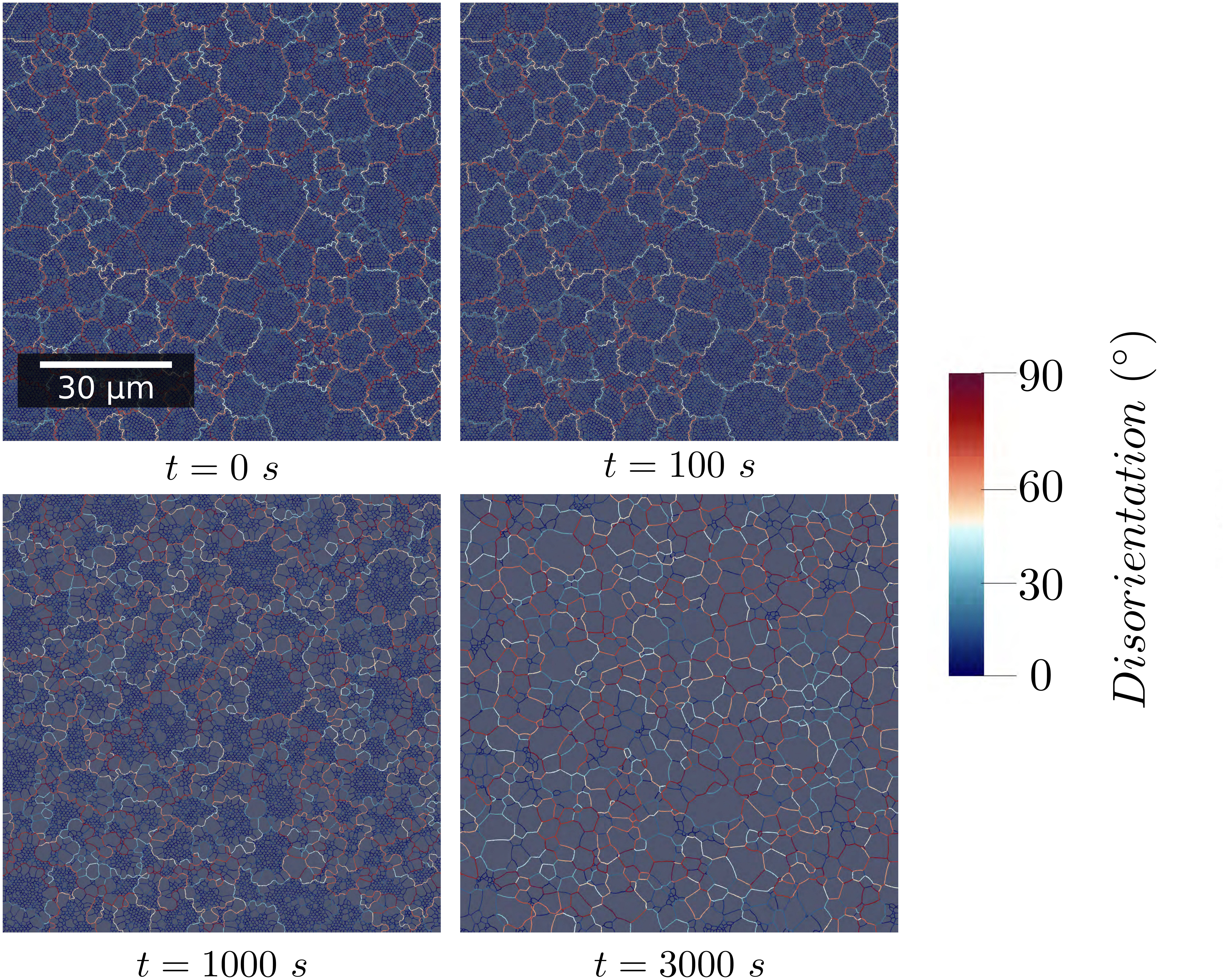}
        \caption{\label{fig:GG_HetGammaHetMobUniformSubgrainSize_FullRVE} Uniform subgrain size distribution - $\gamma_h, ~ M_h$.}
    \end{subfigure}
\caption{Evolution of GB network with time for the four test cases.}
\label{fig:EvolutionGBnetworkGG_full}
\end{figure}

\clearpage

\section{Description of the algorithm underlying the numerical framework}

\begin{figure}[h!]
	\centering
 	\includegraphics[width=0.3\textwidth]{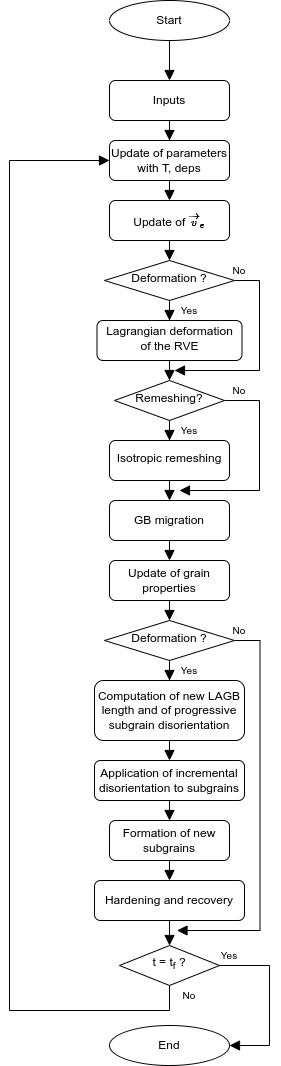}
  \caption{Algorithm describing the numerical steps to simulate microstructure evolutions during hot deformation and holding at temperature.}\label{fig:CDRXalgo}
\end{figure}

\clearpage

\bibliographystyle{unsrtnat}
\bibliography{manuscript}  






\end{document}